\documentclass[%
 reprint,
 superscriptaddress,
nofootinbib,
 amsmath,
 amssymb,
]{revtex4-2}

\usepackage{graphicx}% Include figure files
\usepackage{dcolumn}% Align table columns on decimal point
\usepackage{bm}% bold math
\usepackage[caption=false]{subfig}
\usepackage{siunitx}
\usepackage{xcolor}

\newcommand{\mt}{\si{\micro\tesla}}
\newcommand{\fth}{\si{\femto\tesla\per\sqrt{\hertz}}}
\newcommand{\pah}{\si{\pico\ampere\per\sqrt{\hertz}}}
\newcommand{\mh}{\si{\milli\hertz\per\sqrt{\hertz}}}
\newcommand{\cels}{\si{\degreeCelsius}}

\begin{document}

%\preprint{APS/123-QED}

\title{Optical Magnetometry based Current Source}

\author{Peter A. Koss} 
\email{peter.koss@ipm.fraunhofer.de}
\affiliation{% 
Fraunhofer Institute for Physical Measurement Techniques IPM, 79110 Freiburg, Germany}%
\affiliation{
Instituut voor Kern- en Stralingsfysica, University of Leuven, B-3001 Leuven, Belgium}%

\author{Reza Tavakoli Dinani}
\email{reza.tavakolidinani@kuleuven.be}
\thanks{\newline Peter Koss and Reza Tavakoli Dinani have contributed equally to this work.}
\affiliation{
Instituut voor Kern- en Stralingsfysica, University of Leuven, B-3001 Leuven, Belgium}%

\author{Luc Bienstman}
\affiliation{
Faculty of Engineering Technology, University of Leuven, B-3001 Leuven, Belgium}%

\author{Georg Bison}
\affiliation{Paul Scherrer Institute, 5232 Villigen, Switzerland}

\author{Nathal Severijns}
\affiliation{
Instituut voor Kern- en Stralingsfysica, University of Leuven, B-3001 Leuven, Belgium}%

\date{\today}% It is always \today, today,

\begin{abstract}
We present a current monitoring system based on optical magnetometry, which is able to discriminate true current variations from environmental effects at room temperature. 
The system consists of a dedicated thermally stable magnetic field confining coil and an array of four optically pumped magnetometers arranged in a two-dimensional gradiometer configuration. 
These magnetometers monitor magnetic field variations inside the coil, which correlate to the variations of the driving current of the coil. 
The system uses a digital signal-processing unit to extract and record in real time the magnetic field values measured by the magnetometers, which allows a real time monitoring of the current. 
The system's coil, which is made out of printed circuit boards, can easily be changed to adapt the current-to-field conversion.
Thus, we can expand the applicability of this system to a wide range of currents.
By using this system to actively feedback control a current source we stabilized a current of 20 mA on a level better than $5 \times 10^{-9}$.
\end{abstract}

\maketitle

%\tableofcontents

\section{Introduction}
\begin{figure*}[t!]
	\includegraphics[width=0.8\textwidth]{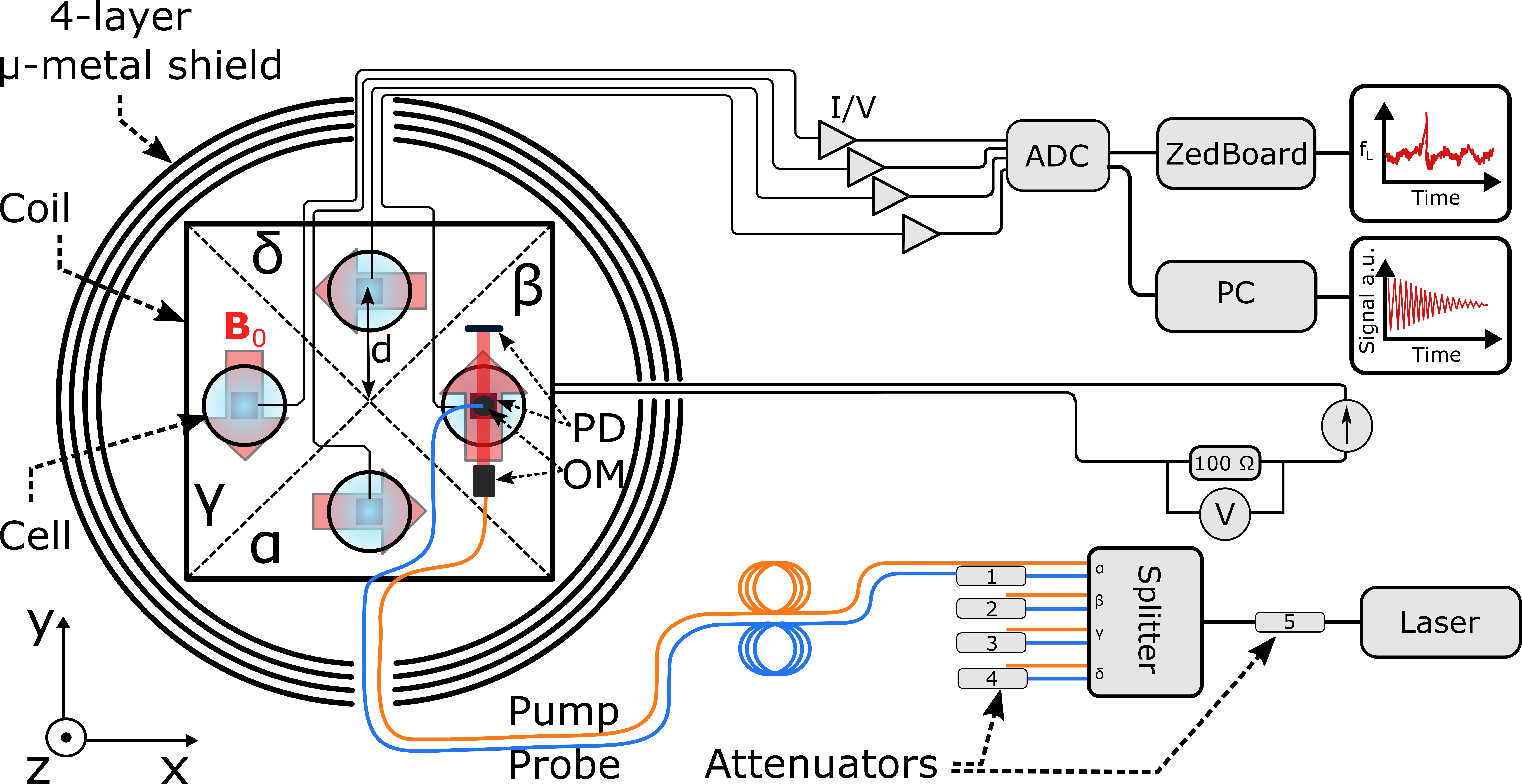}
	\caption{Overview of the current monitoring system (not to scale). Each of the four magnetometers is placed at the center of one of the four quadrants ($\alpha, \beta, \gamma, \delta$) of the field confining coil. This coil stands inside of a 4-layer cylindrical magnetic shield. The magnetometers measure the magnitude of the static magnetic field $|$B$_{j}|$ generated by the coil in quadrant j. The red arrows show the direction of \textbf{B}$_j$. The optical part of the system contains a laser, attenuators, an eight-way beam splitter, optical fibers, optical modules (OM) and photodiodes (PD). It allows us to pump and probe all four magnetometers separately and simultaneously.}
	\label{fig:schematic}
\end{figure*}
\indent The system discussed in this paper was developed for an experiment searching for the permanent electric dipole moment of the neutron (nEDM) \cite{abel2019n2edm, abel2020measurement}.
The spectrometer, located at the Paul Scherrer Institute (PSI) in Switzerland, uses the Ramsey method of time separated oscillating fields~\cite{ramsey1950molecular}. 
This method requires a very stable and low noise magnetic field during its measurement cycles. 
In the PSI experiment, a coil powered by a custom-made current source generates this magnetic field.
The stability and noise level in the measurement volume of the experiment is ultimately limited by the shield if external disturbances dominate. 
Otherwise, it is limited by the stability of this current source.
Furthermore, the control of systematic effects is paramount in this type of experiment~\cite{pendlebury2015revised}. 
The apparatus at PSI has several magnetometry systems, which monitor variations of the magnetic field in situ~\cite{abel2020optically,ban2018demonstration}.
However, we would like to have additional information on the main magnetic field of the experiment, which we intend to get by monitoring the current generating the main magnetic field.\\ 
\indent An ultra-stable current source, which can generate and monitor the currents used for fundamental physics searches, can be built in many different ways. 
The most elementary current source uses a constant voltage source and a precision resistor.
Such a simple design can be improved, e.g., using specially designed operational amplifiers and resistor networks~\cite{drung2015ultrastable}.
However, this rather straightforward path heavily relies on selecting the highest quality electronics components available.
Currently, the best possible stability of a current in the milliampere range was achieved by using a Josephson voltage standard as an external reference to a current source.
Such a system has achieved a stability of a few parts in 10$^{-9}$ for a 50 mA current~\cite{fan2018externally}.
However, the operation of such a system becomes more cumbersome and expensive by the Josephson voltage standard, which only operates with liquid helium, i.e., cooling it down to 4 K~\cite{hamilton2000josephson}.
Another path toward ultra-stable currents is the generation of the current at the single electron level~\cite{pekola2013single}. 
This type of current source is particularly interesting since it strongly suppresses the shot noise on the generated current by using feedback control~\cite{wagner2017strong}.
The technology behind these ``single electron" current sources is very promising; however, they can presently only be used with low currents, i.e., 1 nanoampere or less~\cite{kaneko2016review}.
On the opposite side, with currents of a few ampere, an approach to generate a stable current is using nuclear magnetic resonance (NMR). 
In NMR experiments, such ultra-stable currents were used for a long time~\cite{sasaki1986high,kim1993nuclear}. 
This was made possible by the so-called field-frequency locking technique~\cite{baker1957high}. 
There, a magnetometer lies in the measurement volume of the NMR apparatus.
The reading of this magnetometer serves as the input to a feedback control system. 
Thus, the field of the NMR experiment is stabilized to the limits in sensitivity of the magnetometer.\\
\indent In this paper we propose a method, which is similar to the field-frequency locking of NMR, but monitors the magnetic field produced by the current outside of the measurement volume of the main experiment.
We investigated the sensitivity of a current monitoring system based on optically pumped magnetometers (OPM). 
The main motivation for this work lies with the high sensitivity of OPM in measuring magnetic fields and to convert this attribute to a high sensitivity in measuring currents~\cite{shifrin1996atomic, li2020current}.
To achieve this goal, the system is based on a specially designed magnetic field confining coil, which is made out of printed circuit boards
This allows the path of wires and the coil constant to be adapted to a specific application~\cite{koss2017pcb, crawford2020physical}.
This coil contains an array of four OPM, each placed in a different section of the coil, which monitor the local field generated by the current passing through it. 
In this work, we used cesium (Cs) and potassium (K) based OPM and compared their performance for our application.
This system, used with a feedback control, can generate an ultra-stable current.
We present results of an actively stabilized current of 20 mA which shows a stability better than $5 \times 10^{-9}$.\\
\section{Experimental Setup}
The system described in this paper, which is shown schematically in Fig.~\ref{fig:schematic}, confines four Cs or K magnetometer modules inside of a field-confining coil. 
A mu-metal shield with four concentric cylindrical layers isolates the magnetometers from external electromagnetic interference. 
A commercial current source drives the current in the coil.
This generates a circumferential static magnetic field \textbf{B}$_0$ such that each opposing magnetometer pair senses an anti-parallel \textbf{B}$_{0}$ as it is shown in Figs.\,\ref{fig:schematic} and \ref{fig:V2}. \\
\indent Each magnetometer has its own RF coil, pump and probe laser beam, which allows us to operate them in an RF-pulsed mode (see Fig. \ref{fig:RF_magnetometer}).
The circularly polarized pump beam orients the atomic spins along the local magnetic field \textbf{B}$_j$.
The spin precession of the alkali atoms at the Larmor frequency $f_{L}$ is monitored using a low power circularly polarized probe beam. 
The signal provided by the magnetometers is a photodiode current, which converts to a voltage signal via a transimpedance amplifier before being digitized by an analog-to-digital converter (ADC).\\
\indent The Larmor frequency $f_{L}$ can be extracted from the digitized signal in real time, using a digital signal processing unit (DSP), which we call the online mode of analysis.
Alternatively, $f_L$ can be extracted in an offline mode in which signals are stored on a computer, and then fitted with an exponentially decaying sinusoid function. 
We call this, the offline mode of analysis. 
Either method yields Larmor frequencies $f_L$, which are used to infer $i$, the current passing through the coil:
\begin{equation}
i = \frac{2 \pi}{\gamma_F \lambda} f_{L}~,
\end{equation}
where $\lambda$ is the coil constant and $\gamma_F$ is the gyromagnetic ratio of the hyperfine ground state level of the respective alkali atoms.
The gyromagnetic ratio of the ground state in alkali atoms with electronic spin $J$, nuclear spin $I$ and atomic spin $F$ can be derived from the ground state’s electronic ($g_J$) and nuclear ($g_I$) g-factors~\cite{arimondo1977experimental} using
\begin{equation}
    \begin{aligned}
     g_F &= \frac{g_J}{2} \frac{F(F+1) + J(J+1) - I(I+1)}{F(F+1)} \\
         &+ \frac{g_I}{2} \frac{F(F+1) - J(J+1) + I(I+1)}{F(F+1)}~,
    \end{aligned}
\end{equation}
and
\begin{equation}
    \frac{\gamma_F}{2 \pi} = g_F \frac{\mu_B}{h}~,
\end{equation}
where $\mu_B$ is the Bohr magneton, $h$ is Planck's constant and $g_F$ is the atomic g-factor.
For the Cs ground state $6^{2}S_{1/2}$ with spin F = 4 the gyromagnetic ratio is $\gamma_4 / 2\pi$ = 3.49862095(35) Hz/nT.
In a similar way, we use the $4^{2}S_{1/2}$ K ground state with spin F = 2 whose gyromagnetic ratio is $\gamma_2 / 2 \pi$ =~{7.00466013(84)}~Hz/nT.
Thus, this is an attempt to link a current measurement to atomic constants.
\subsection{Field Confining Coil}
\indent The requirements on the coil for this current monitoring concept link to the type of magnetometry we are using.
The coil has to produce a very uniform magnetic field, as gradients broaden the magnetometer's magnetic resonance lines.
This broadening degrades the sensitivity on the magnetic field reading~\cite{pustelny2006influence}.
The coil also has to confine the field within itself such that the stray field is not affected by the environment. 
The most important criterion for our application is the gradiometer configuration of the magnetometers, as we want to suppress environmental magnetic field perturbations. 
The field direction inside of the coil should be circumferential.
This allows us to have anti-parallel magnetic fields for pairs of magnetometers sensitive along the $x$ and $y$ directions (see Fig.~\ref{fig:schematic}). 
We have a total of four magnetometers, each installed in a separate quadrant of the coil. 
In the presence of a coil driving current $i(t)$ and an external perturbing magnetic field $\mathbf{B}_\mathrm{ext}(x, y, t)$, where we assumed $\mathbf{B}_\mathrm{ext}$ is a function of time and the coordinates $(x, y)$, the magnetic field measured by the magnetometers at a time $t$ is:
\begin{equation}
    \mathbf{B}_\mathrm{j} (x, y, t) = \mathbf{B}_{\mathrm{coil, j}} (x, y, t) + \mathbf{B}_{\mathrm{ext, j}} (x, y, t)
\end{equation}
where the index stands for the place of the module j = $\alpha, \beta, \gamma, \delta$.\\
\indent Taking the simplified case where $\mathbf{B}_\mathrm{ext}$ has constant gradients along x and y we get:
\begin{equation}
\mathrm{Pair~1} =
\begin{cases}
    \text{B}_{\alpha} \simeq i \lambda_{\alpha}  + (\text{B}_\mathrm{ext, 0} - d\text{B}^{\prime}_{y})~,\\
    \text{B}_{\delta} \simeq i \lambda_{\delta} - (\text{B}_\mathrm{ext, 0} + d \text{B}^{\prime}_{y})~,\\
\end{cases}
\label{eq:pair1}
\end{equation}
\begin{equation}
\mathrm{Pair~2} =
\begin{cases}
    \text{B}_{\beta} \simeq i \lambda_{\beta}  + (\text{B}_\mathrm{ext, 0} + d\text{B}^{\prime}_{x})~,\\
    \text{B}_{\gamma} \simeq i \lambda_{\gamma} - (\text{B}_\mathrm{ext, 0} - d\text{B}^{\prime}_{x})~,\\
\end{cases}
\label{eq:pair2}
\end{equation}
where we grouped the sensors by opposing gradiometer pairs.
$\text{B}_\mathrm{ext, 0}$ is the magnitude of the perturbing field at the origin. 
$\text{B}^{\prime}_{x} = \frac{\partial \text{B}_\mathrm{ext}}{\partial x}(0, 0)$ and $\text{B}^{\prime}_{y} = \frac{\partial \text{B}_\mathrm{ext}}{\partial y}(0, 0)$ are the field gradients at the origin. 
The parameter $d$ is the distance of the center of the magnetometer's vapor cell from the origin of the coordinate system, which is located at the center of the coil.
The factors $\lambda_{\alpha}, \lambda_{\beta}, \lambda_{\gamma}$ and $ \lambda_{\delta}$ are the coil constants at the locations of the magnetometers.
They can be measured accurately.\\
\indent Using Eqs. \ref{eq:pair1} and \ref{eq:pair2}, the driving current of the coil reads:
\begin{equation}
i(t) = \frac{ \sum_{j}\text{B}_{j} + 2d (\text{B}^{\prime}_{x} - \text{B}^{\prime}_{y}) }{\sum_{j}\lambda_{j} }~, \label{eq_I}
\end{equation} 
which shows that this gradiometer configuration can suppress a uniform \textbf{B}$_\mathrm{ext}$.
Higher coil constants and smaller dimensions of the coil suppress the influence of the constant gradients.
However, for higher order non-uniform cases, we require more information about the first order gradient of \textbf{B}$_\mathrm{ext}$ along the $x$- and $y$-axes. 
This can be achieved by installing a third magnetometer along these axes. 
For the investigation in this paper, we limit ourselves to only two magnetometers per axis.\\
\indent The maximum contribution of \textbf{B}$_\mathrm{ext}$ to a current measurement of one sensor, including the first order gradient, is $(\text{B}_\mathrm{ext, 0} + d\text{B}^{\prime})/\lambda$. 
Using Eq.\,(\ref{eq_I}) and assuming the same coil constant $\lambda$ and the same magnitude of gradients for all quadrants; this contribution is $d\text{B}^{\prime}/2\lambda$.
One can use the ratio of these contributions to define a shielding factor (SF) associated with the gradiometer configuration of the coil.
This defines to what extent the effect of \textbf{B}$_\mathrm{ext}$ on the measured current is suppressed. 
So, 
\begin{equation}
    \mathrm{SF} = 2(\text{B}_\mathrm{ext, 0}/d \text{B}^{\prime} + 1)~. \label{eq: SF}
\end{equation}
The factor 2 in SF comes from the two pairs of magnetometers. 
This factor is unity when we only use one pair of magnetometers. 
Thus, Eq.\,(\ref{eq: SF}) shows how a two-dimensional gradiometer configuration can outperform a one-dimensional configuration. 
To improve SF it is clear that $d$ should be reduced.\\
\begin{figure}[t!]
	\includegraphics[width=0.48\textwidth]{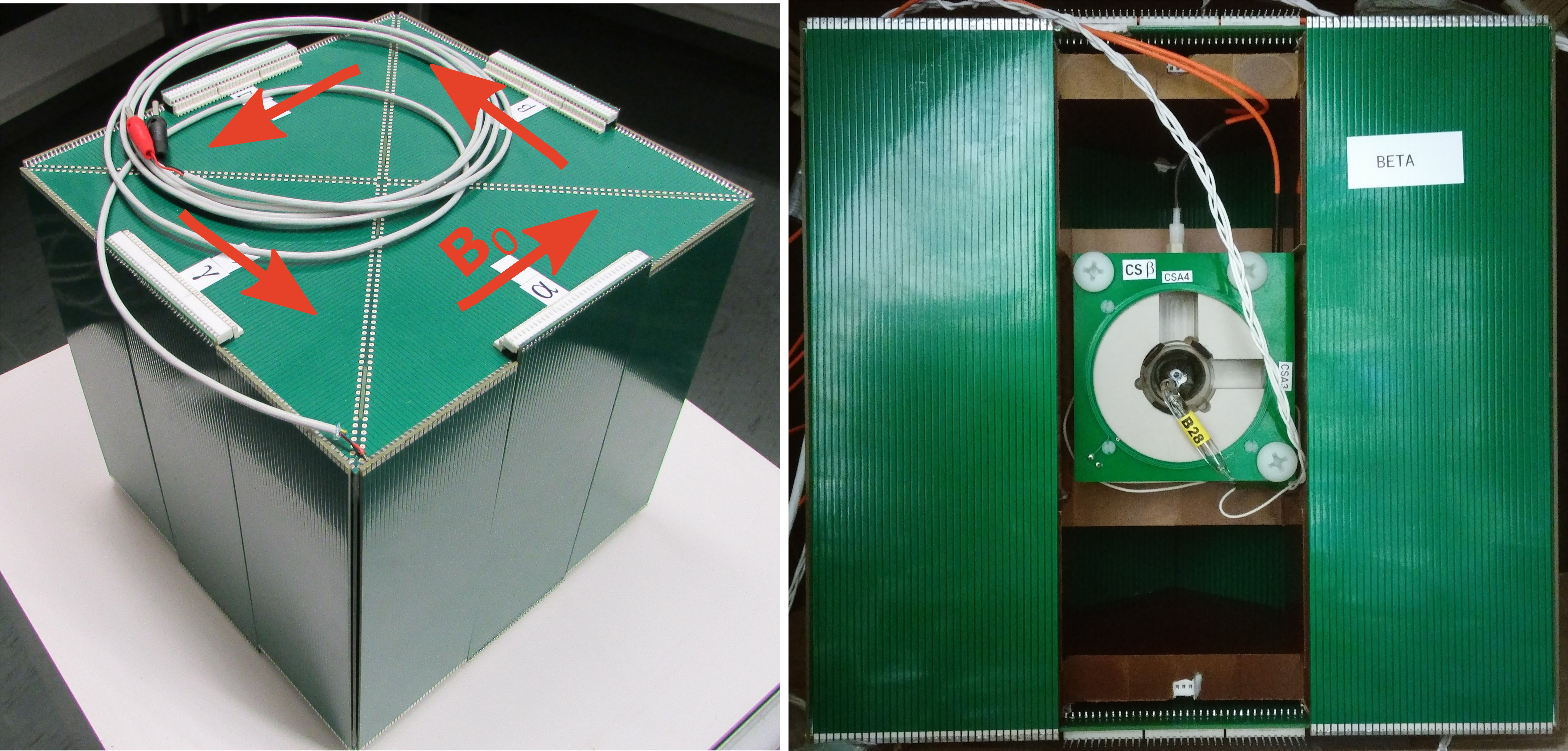}
	\caption{Pictures of the field confining coil used in this work. It is entirely made out of printed circuit boards (PCB). On the left, the full coil is shown with $\alpha$, $\beta$, $\gamma$ and $\delta$ quadrants, which open via non-magnetic PCB headers: long white connectors. The arrows on the top panel show the direction of the magnetic field when applying a positive current. On the right, an open quadrant with the mount and a Cs magnetometer is shown. The fibers and cables used for the magnetometer leave the coil through a hole in the top. Temperature sensors (DS18B20) were glued into the mount on the top and bottom: white spots in the mount.}
	\label{fig:V2}
\end{figure}
\indent The coil used in this work is shown in Fig.~\ref{fig:V2}.
It is entirely made out of multilayered printed circuit boards (PCB). 
Each quadrant has 100 equidistant current loops.
The magnetic field is perpendicular to those current loops and points into different directions in each quadrant.
The four quadrants open separately with the use of removable front panels.
These front panels are connected to the body of the coil with non-magnetic PCB headers (females: molex~KK~4455 series, males: molex~5046~series).
Each one of these quadrants contains a mount made of PFCC201, a non-magnetic and low thermal expansion material.
These mounts are glued into the quadrant using silicon glue.
An OPM can be mounted inside each quadrant as shown in Fig.~\ref{fig:V2}. 
The total resistance of the coil is R~=~58~$\Omega$ and the design coil constant is 0.5~\mt/mA. 
The characteristics of the coil design used in this work are similar to a coil design presented before~\cite{koss2017pcb}.\\
\indent The coil must be robust against temperature fluctuations, as this can affect all four quadrants in the same way; thus generating a fake current drift.
A simple aluminum thermal chamber, which was heated with silicon oil, was used for all measurements associated with K magnetometers and kept the operating temperature at 330 K.
All measurements associated with Cs magnetometers were performed at room temperature.
The aluminum chamber accommodates the coil.
In order to fit this assembly inside the innermost layer of the magnetic shield, a volume of 40$\times$40$\times$40~cm$^3$ is required. 
This setup, allowed us to make dedicated measurements of the coil constant as a function of temperature.
We determined that a relative change in the coil constant is less than $4 \times 10^{-5}$ K$^{-1}$. 
At room temperature we measured a temperature stability of 0.1 mK after 2~min and 10 mK after 1~h. 
Thus, with this thermally stable coil constant we do not expect any perturbing temperature effects on the relevant time scales. 
\subsection{Magnetometers}
\begin{figure}[h!]
	\includegraphics[width=0.45\textwidth]{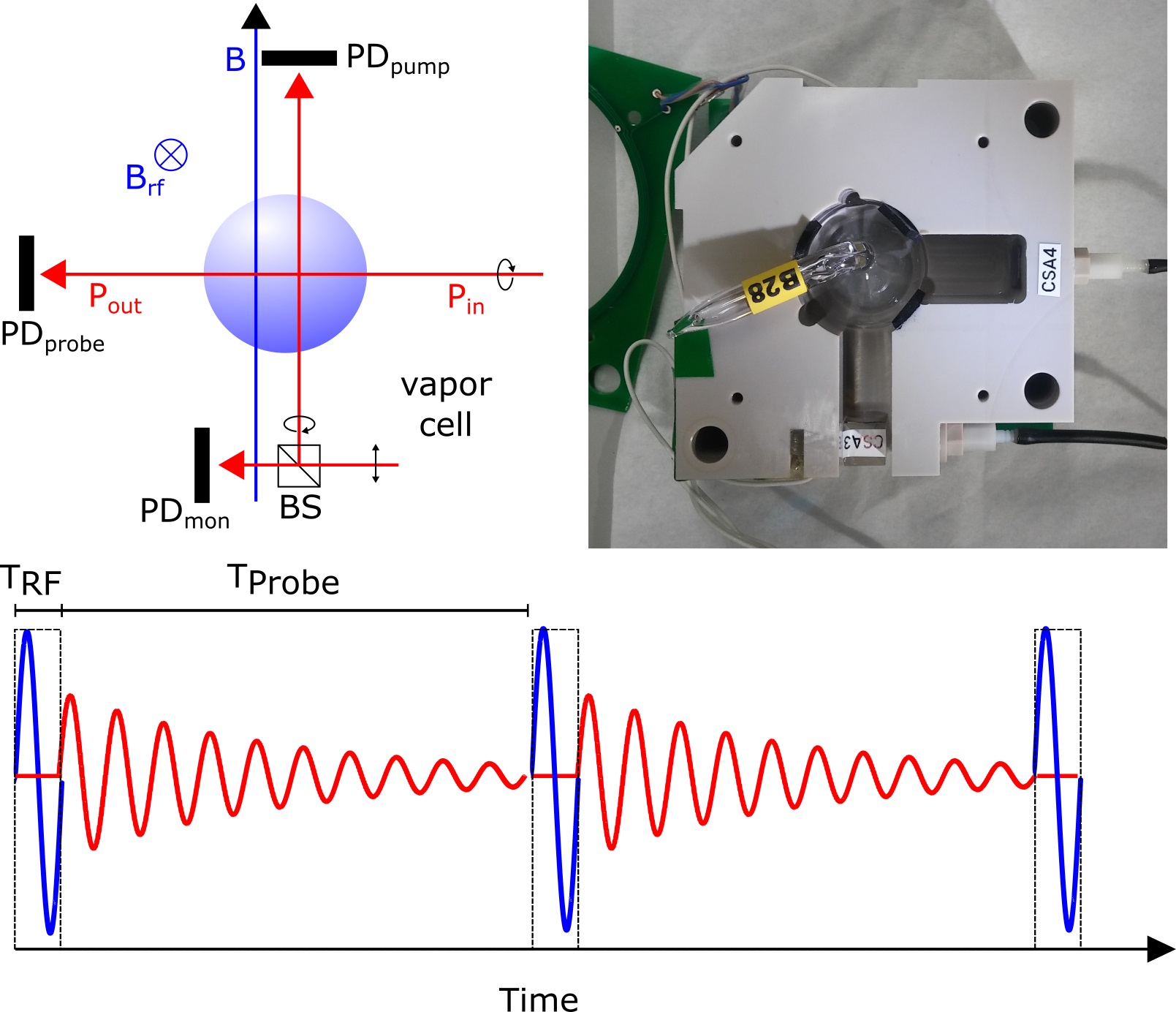}
	\caption{Schematic and picture of the optically pumped magnetometer (OPM) used for this work. The top left schematic represents an OPM module, where the red arrows represent the laser beams, the three photodiodes are labeled PD and BS is a beam splitter. The pump beam is parallel to the applied magnetic field and the probe beam is perpendicular to the pump beam. The top right picture shows a Cs OPM, whose frame is made out of PEEK. The vapor cell (labeled B28) is paraffin coated. CSA3 and CSA4 are circularly polarizing optics. The green PCB is one of two coils, which arrange like Helmholtz coils. The bottom schematic shows the operation pulse sequence for these OPM. A short, typically 1-3 cycle long RF-pulse is applied on the spin-polarized medium of the sensor: the blue pulse. This initiates the free spin precession, which the probe beam monitors: the red signal. Note that the frequencies and amplitudes are not to scale for clarity.}
\label{fig:RF_magnetometer}
\end{figure}
\begin{figure}[t!]
    \subfloat[\label{subfig:Cs_sens}]{%
    \includegraphics[width=0.5\columnwidth]{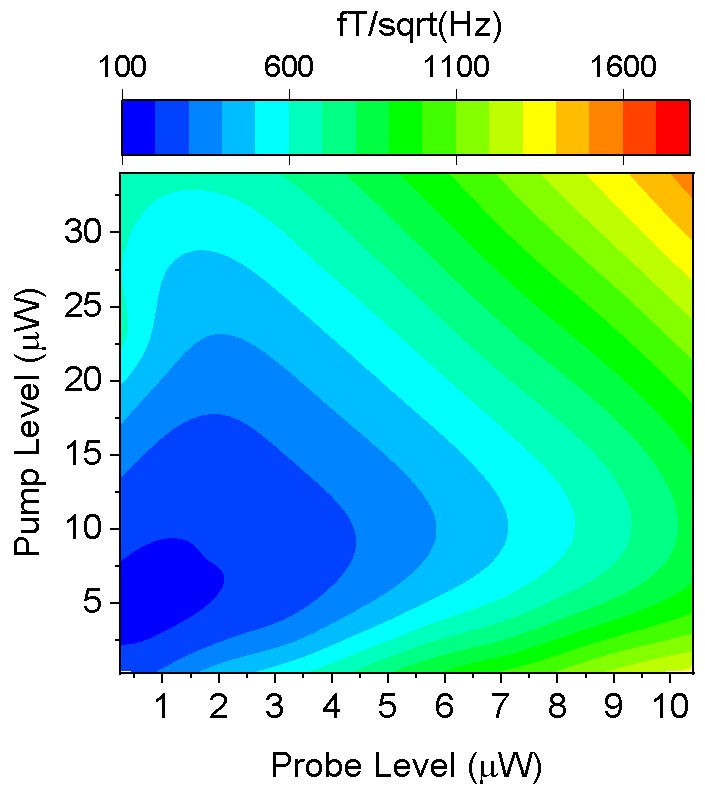}%
    }
    \subfloat[\label{subfig:K_sens}]{%
    \includegraphics[width=0.5\columnwidth]{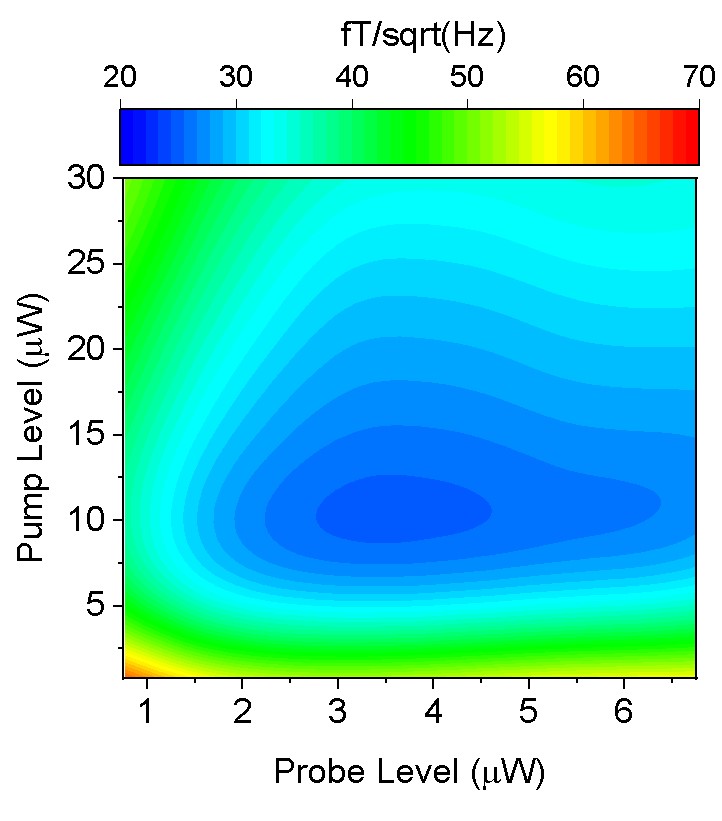}%
    }
	\caption{Shot-noise limited sensitivity maps as a function of the laser power at 1 \mt~for Cs and K. (a) Cs magnetometer held at room temperature with a vapor cell diameter of 30~mm. (b) K magnetometer held at 50\cels~with a vapor cell diameter of 70~mm. The experimentally determined maximum sensitivity lies at 180~\fth~for Cs and 20~\fth~for K magnetometers. In both cases the FSP signal duration was 70~ms.}
	\label{fig:Cs_K_sens}
\end{figure}
\begin{figure}[t]
	\includegraphics[width=0.47\textwidth]{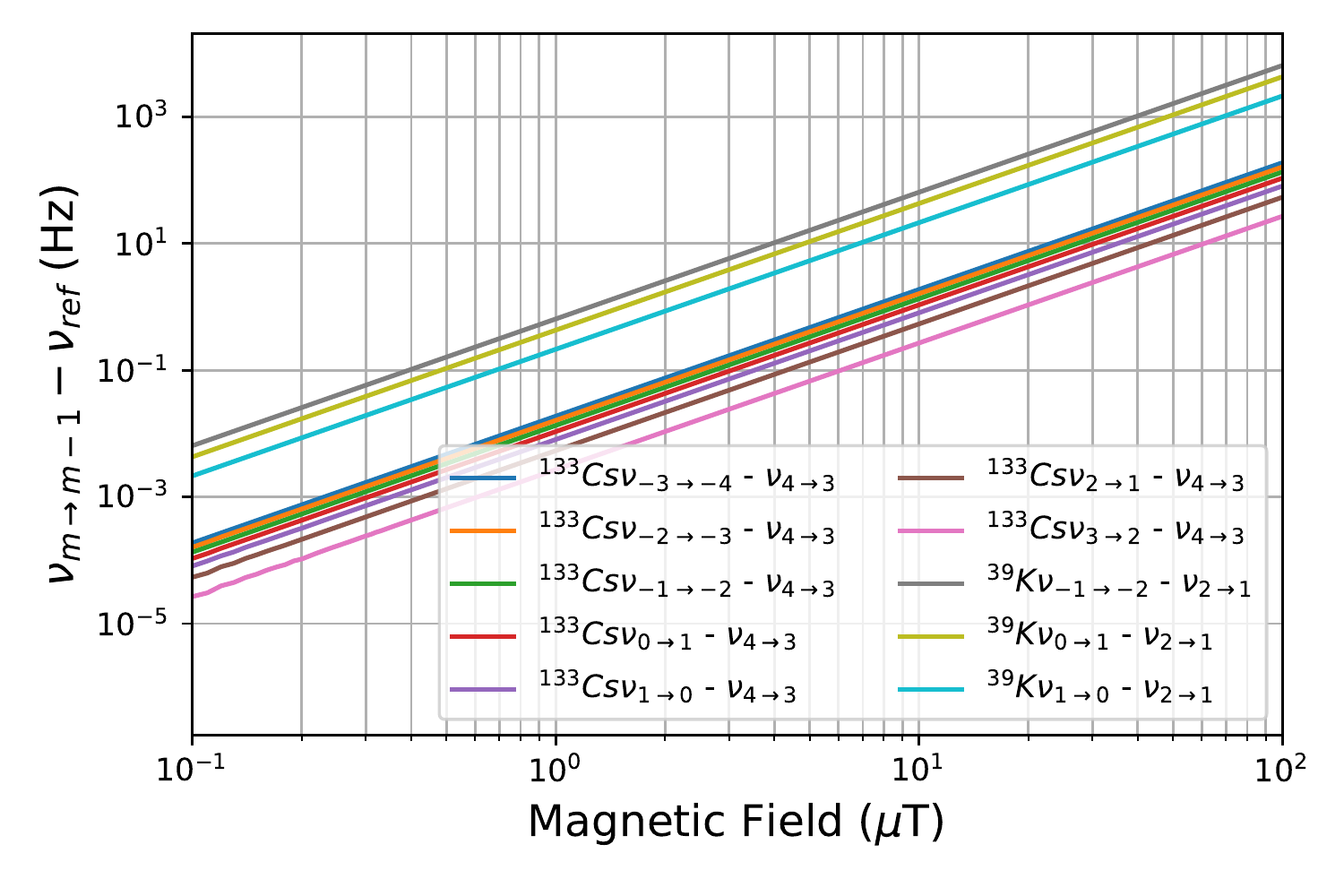}
	\caption{Non-linear Zeeman effect in potassium ($^{39}$K) with F\,=\,2 and cesium ($^{133}$Cs) with F\,=\,4, i.e., non-linear splitting of the magnetic sub-levels, written m. The large linear contribution of the Zeeman effect was subtracted by using one sublevel as a reference ($\nu_{4 \rightarrow 3}$ and $\nu_{2 \rightarrow 1}$). The effect is much larger in the case of potassium due to a smaller hyperfine splitting of the element.}
	\label{fig:nl}
\end{figure}

The magnetometers used in this work were optically pumped magnetometers.
They are based on the detection of a magnetic resonance signal in a spin-polarized vapor of alkali metals~\cite{bell1957optical}.
Their relative simplicity of use and very high sensitivity have made them a very popular tool in fundamental physics searches~\cite{alexandrov1992optically,budker2007optical}.
Figure~\ref{fig:RF_magnetometer} shows the magnetometer modules used for this work.
Two different elements were chosen and used in different modules for comparison: Cs and K.
The alkali metal vapor of these modules is contained in a paraffin coated glass cell.
This coating prevents depolarization of the vapor through collisions of the atoms with the glass wall~\cite{robinson1958preservation,castagna2009large}.\\
\indent The laser light used for the optical pumping of the medium was provided by two single mode diode lasers from TOPTICA.
The wavelength of one was stabilized to the F\,=\,2\,$\longrightarrow$\,F$^\prime$\,=\,1 transition of the K $D_1$ line, i.e., at 770~nm.
The other laser was stabilized to the F\,=\,4\,$\longrightarrow$\,F$^\prime$\,=\,3 transition of the Cs $D_1$ line, i.e., at 894~nm.
The stabilization of the laser wavelength was performed with the use of a Doppler-free saturation spectroscopy setup.
The laser light is guided to the module via multi-mode fibers.
At the end of these fibers, a beam collimating optical module (OM) with a linear polarizer is mounted.
Another linear polarizer is part of the circularly polarizing optics (CSA3 and CSA4 in Fig.~\ref{fig:RF_magnetometer}).
By changing the alignment of these polarizers, one can continuously change the laser power passing through the vapor cell of the magnetometer.\\
\indent The OPM presented here use the so-called free spin precession (FSP) mode of operation~\cite{grujic2015sensitive}.
It consists of polarizing the alkali vapor with a pumping beam, then monitoring the free precession of the spin-polarization of the alkali vapor.
This is in contrast with other modes, e.g., the very popular $M_x$ type, which continuously drives the magnetic resonance~\cite{aleksandrov1995laser}.
The advantages of this FSP-mode are much lower systematic effects on the readout of the Larmor frequency, less crosstalk between the RF-fields of the modules and a good sensitivity.
It typically reaches levels lower than 100 \fth~in the shot noise limit~\cite{grujic2015sensitive}.\\
\indent In our case, the FSP is initiated with a tipping $\pi/2$-pulse~ \cite{avrin1989optically,afach2015highly}. 
This pulse is typically 1-3 periods of a resonant RF-pulse of large amplitude.
Since it is very short in time, it is very broad in the frequency domain.
The advantage is that the frequency of the pulse does not have to be very close to the Larmor frequency of the alkali vapor.
This tipping pulse occurs typically every 100 ms, which sets a 10\,Hz sampling rate for the current measurement. 
Higher sampling rates are possible, but this will degrade the sensitivity of the magnetometers. 
The FSP signal recorded by the photodiode is amplified by a transimpedance amplifier, whose nominal gain is $G = 10^6$~V/A.
This voltage signal is digitized using either a 16~bit ADC with a maximum sampling rate of 1\,MS/s or a 24~bit ADC sampling at 48.8~kS/s. 
The high resolution ADC is used in the online mode of analysis, while the lower resolution one is used in the offline mode.\\
\indent The digitized signal is an exponentially decaying sinusoidal time series.
For a given sampling rate $r_S$, with $r_S/2 \geq f$, the recorded voltage reads:
\begin{equation}
    \mathrm{FSP =}
    \begin{cases}
	V(n) &= V_0 \sin (2 \pi f_L n \Delta t + \phi) e^{-\Gamma n \Delta t} + w(n)~,\\
	n &= 0, 1, 2, ..., N - 1~,
	\label{eq:FSP_signal}
	\end{cases}
\end{equation}
where $f_L$ is the Larmor frequency, $\Gamma$ is the relaxation rate imposed by the experimental conditions and $\phi$ is the initial phase.
The signal is perturbed by white noise $w(n)$ and sampled with a sampling interval of $\Delta t = 1/r_S$.
The total recording time is $T_r = N/r_S$.\\
\indent The statistical sensitivity of the magnetometer depends on the precision with which we can extract the frequency from the measured signal.
The statistical tool which gives the lowest variation of an unbiased frequency estimator is the Cram\'er-Rao lower bound~\cite{rao1992information}.
The Cram\'er-Rao lower bound for Eq. \ref{eq:FSP_signal} in terms of experimentally measurable parameters is~\cite{gemmel2010ultra}
\begin{equation}
	\sigma_f = \frac{\sqrt{12}}{(V_0/\rho_V) T_r^{3/2}} \sqrt{C}~,
	\label{eq:crlb}
\end{equation}
where $V_0/\rho_V$ is the signal-to-noise ratio, $T_r$ is the length of the signal and C is given by
\begin{equation}
	C = \frac{N^3}{12} \frac{(1 - z^2)^3 (1 - z^{2N})}{z^2 (1 - z^{2N})^2 - N^2 z^{2N} (1-z^2)^2}~,
	\label{eq:crlb_c}
\end{equation}
where $z = e^{-\Gamma \Delta t}$.
We characterize the sensitivity of the magnetometers in the shot-noise limit.
This means the noise level in Eq.~\ref{eq:crlb} is given by $\rho_V = G \sqrt{2 e I_{DC}}$, where $e$ is the electron charge, $G$ is the gain of the transimpedance amplifier and $I_{DC}$ is the DC part of the photodiode current.\\
\indent The sensitivity of the FSP mode depends on the atomic physics parameters of Cs and K in a non-trivial way~\cite{grujic2015sensitive}. 
For this reason, we scanned the laser powers of the pump and the probe beam in order to map the sensitivity and find the optimal operation point. 
The results of this investigation are shown in Fig.~\ref{fig:Cs_K_sens} for a 1 \mt~holding field.
The sensitivity of the magnetometers is given by the Cram\'er-Rao lower bound in Eq.~\ref{eq:crlb}.
The only parameter affected by the field strength is the linewidth of the signal $\Gamma$.
\begin{figure}[t]
	\includegraphics[width=0.47\textwidth]{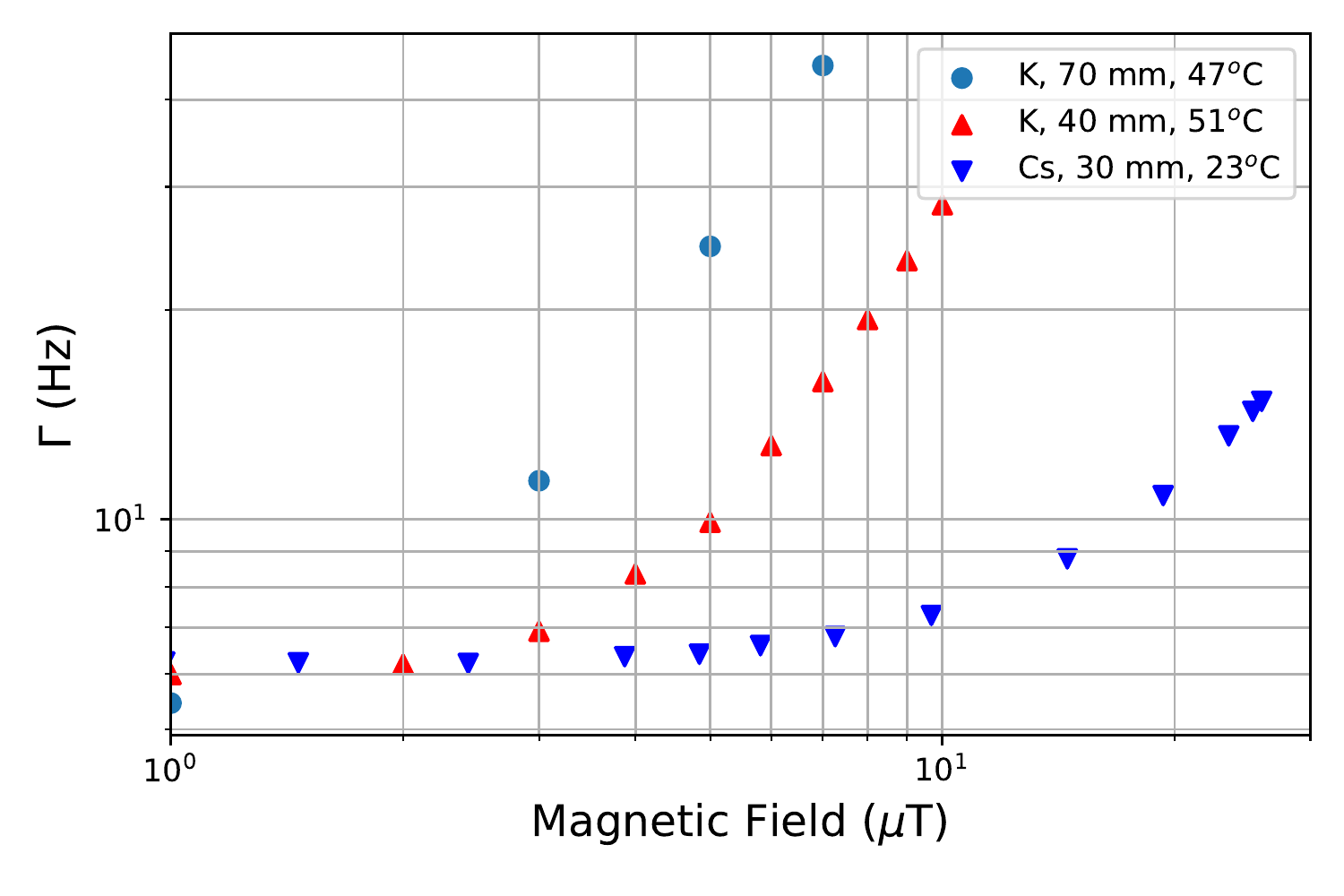}
	\caption{Linewidth of Cs and K free spin precession (FSP) signals as a function of the applied magnetic field strength. The linewidth at 1~\mt~is close to the intrinsic linewidth of the cells \cite{groeger2006high}. At higher fields, the linewidth of the FSP signal broadens due to the non-linear Zeeman effect and magnetic field gradients. The diameter of the cell is indicated in mm for each case.}
	\label{fig:cs_linewidth}
\end{figure}
Figure~\ref{fig:cs_linewidth} shows the dependence of the linewidth of Cs and K magnetometers as a function of the applied magnetic field strength. 
The cell diameter for the Cs magnetometer was 30~mm and for K they were 40~mm and 70~mm. 
Cs magnetometers were used at room temperature (23\cels), but K magnetometers were used at 51 and 47\cels. 
The higher operating temperature for K is necessary because the vapor pressure of K is much lower than the vapor pressure of Cs.
The measurements all use the coil. 
We clearly see how the linewidth is very strongly influenced by the field strength in the case of a 70~mm diameter K vapor cell. 
This behavior can be improved by using a smaller vapor cell, e.g., a 40~mm diameter one. 
This means that K is very sensitive to magnetic field gradients even at low magnetic field strengths~\cite{cates1988relaxation,pustelny2006influence}.
In comparison, Cs is much more robust to magnetic field gradients. 
Other parameters that affect the linewidth broadening can be explained by the non-linear Zeeman effect, which the Breit-Rabi formula describes~\cite{breit1931measurement}. 
This effect is shown on Fig.~\ref{fig:nl} for both Cs and K.
Over the same range of magnetic field strengths, this effect is larger by more than an order of magnitude for K. 
Although the tested K magnetometer was more sensitive than the Cs magnetometer, the magnetic field dependent linewidth broadening limits its sensitivity. 
As our application typically requires currents of 12 mA or more, corresponding to a field of 6 \mt~in the coil, we have decided to use Cs magnetometers for current measurements.\\
\begin{figure}[t!]
	\includegraphics[width=0.5
\textwidth]{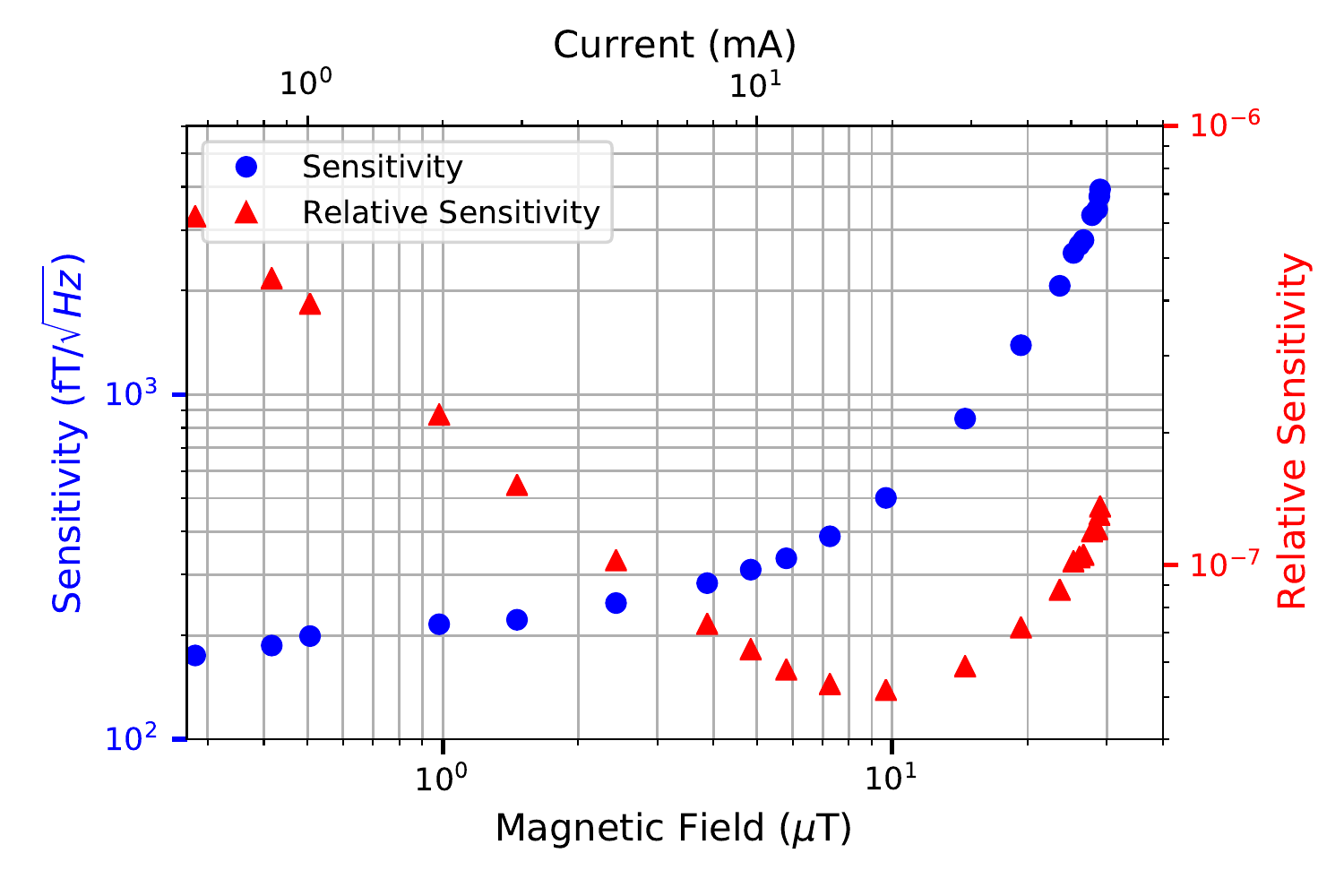}
	\caption{Shot noise limited ({\large \color{blue}{$\bullet$}}) and relative ({\color{red}{$\blacktriangle$}}) sensitivities of Cs magnetometers used in the system as a function of magnetic field and current. The relative sensitivity is shown for a 1~Hz~bandwidth. The sensitivity is limited at high magnetic fields due to Cs line broadening.}
	\label{fig:sensitivity_vs_field}
\end{figure}
\indent Figure~\ref{fig:sensitivity_vs_field} shows the shot noise limited ($\sigma_\mathrm{SN}$) and relative ($\sigma_\mathrm{Rel}$) sensitivities of the Cs magnetometers as a function of the magnetic field in the coil.
At low fields, where the linewidth broadening is small, $\sigma_\mathrm{SN}$ worsens moderately. 
However, due to linewidth broadening, $\sigma_\mathrm{SN}$ gets worse at higher fields.
The relative sensitivity improves up to 10 \mt, where we see $\sigma_\mathrm{Rel} = 5 \times 10^{-8}$.
This corresponds to a sensitivity of 1\,nA on a 20\,mA current applied to the coil.\\
\subsection{Digital Signal Processing}
\begin{figure}[t!]
	\includegraphics[width=0.48\textwidth]{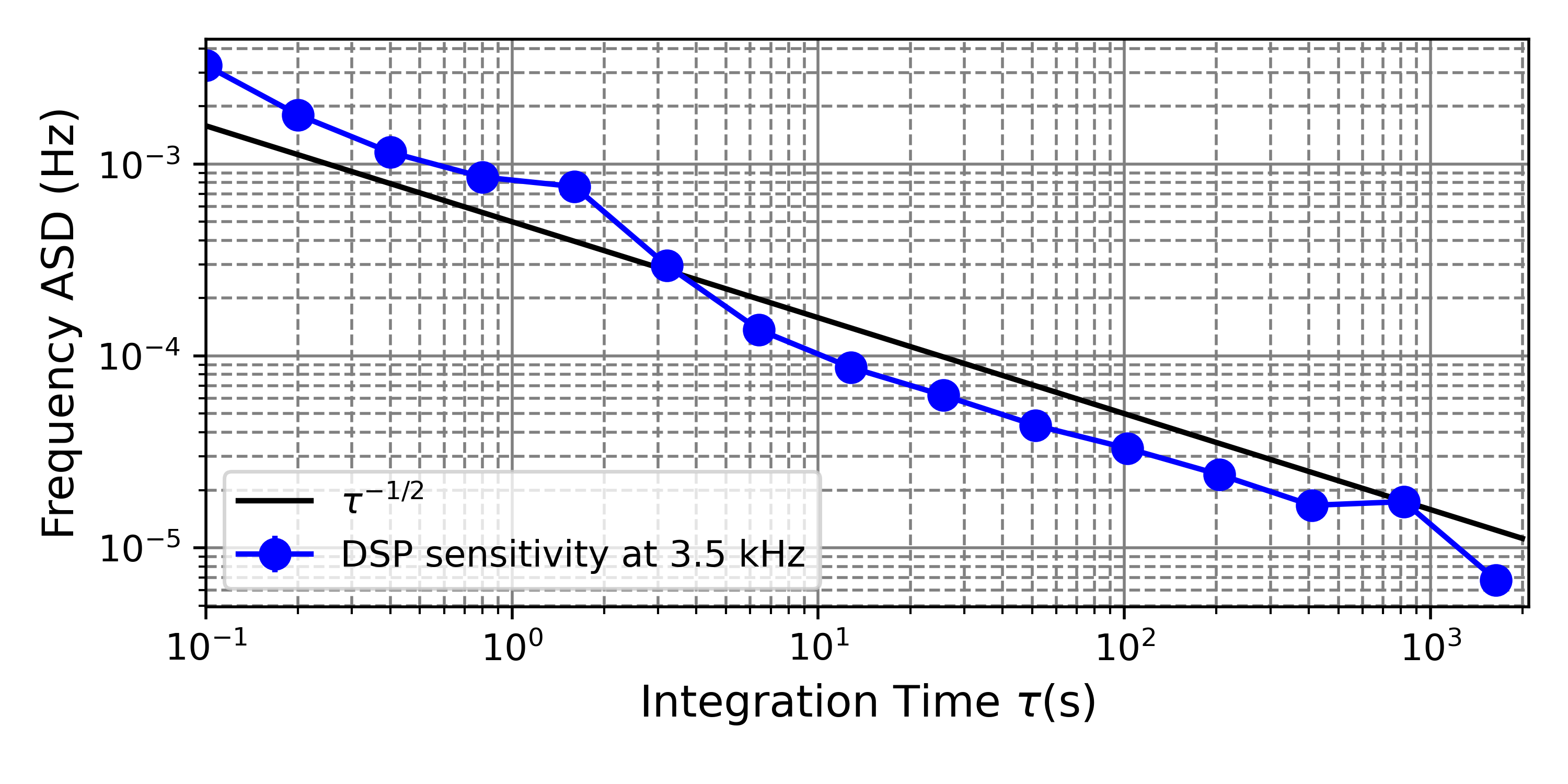}
	\caption{Allan Standard Deviation (ASD) of the measured frequency of a continuous test sinusoidal signal using our DSP system at 3.5\,kHz. At an integration time of 0.5 s it has a sensitivity of 1~mHz on the 3.5 kHz. The lines between the points are merely a guide for the eye.}
\label{fig:DSP-evaluation}
\end{figure}

\indent We developed a DSP system to extract the precession frequency $f_{L}$ of Cs atoms in all four magnetometers simultaneously and in real time. 
The core of the system is a ZedBoard \cite{Zedboard}, which is a prototyping board featuring a highly integrated system on a chip (SoC).
The input of the DSP is a 24-bit ADC with a sampling rate of 48 kS/s. 
This DSP system can accept up to eight input channels, where the input signal is an amplified analog FSP voltage signal.\\
\indent The digitized signal is filtered, amplified and then demodulated, using a lock-in algorithm, to extract the instantaneous phase $\theta$(t) of the signal compared to a local oscillator. 
By knowing the accumulated phase $\theta$(t) as a function of time, one can find the frequency $f_{L}$ using a linear regression. 
To keep track of $f_{L}$ over time, a proportional-integral-differential (PID) algorithm was used. 
The associated error signal in the PID algorithm is the difference between the measured frequency and the presumed stable local oscillator frequency. 
Both the demodulation and PID algorithms are implemented on the field-programmable gate array (FPGA) of the ZedBoard. 
The linear regression of the phase as a function of time was implemented on the processor of the ZedBoard.
This lets us calculate the Larmor frequency $f_{L}$ of a typically 70\,ms long FSP signal from all four modules in less than 30\,ms. \\
\indent We evaluated the sensitivity of this DSP system at 3.5~kHz using a continuous sinusoidal test signal generated at an amplitude comparable to the average rms value of a standard FSP signal. 
The test signal was generated by a Zurich Instrument lock-in amplifier model MLFI. 
Figure~\ref{fig:DSP-evaluation} shows the modified Allan standard deviation (ASD) of the measured 3.5 kHz frequency as a function of the integration time $\tau$. 
All Allan plots in this text use the so-called modified Allan standard deviation as it yields much better confidence intervals at high integration times~\cite{riley2008handbook}.
The sensitivity of the DSP system is limited by pure white noise as it is clear from the straight line with slope $\tau^{-1/2}$. 
One expects to measure the frequency of a clean test signal with a sensitivity of about 1~\mh~using our DSP system.\\
\begin{figure}[t!]
	\includegraphics[width=0.48\textwidth]{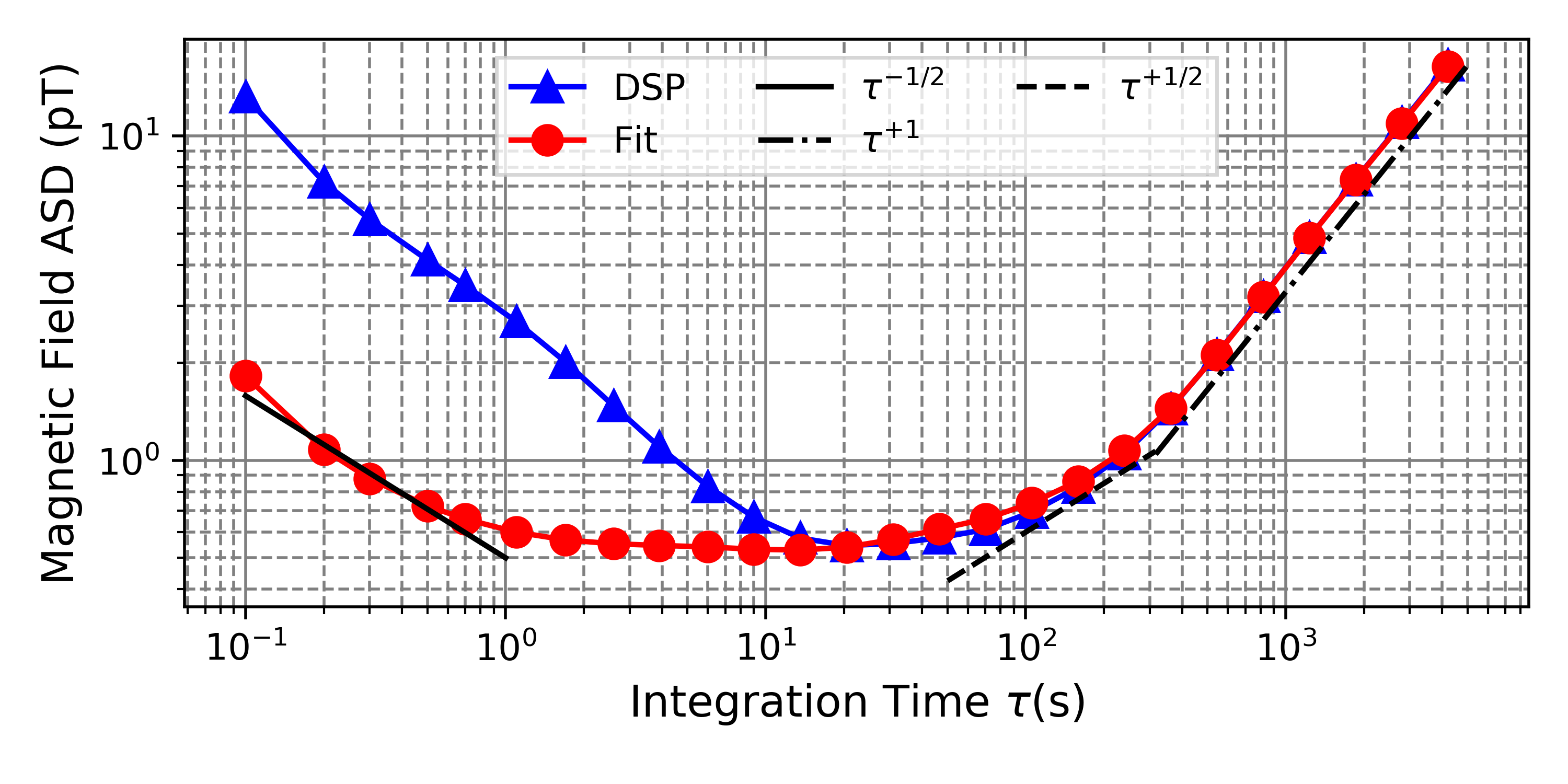}
	\caption{ASD plot of the measured magnetic field inside of our coil using Cs magnetometers at 1~\mt~and at room temperature. The FSP signals of the magnetometers are analyzed using the DSP system ({\color{blue}{-$\blacktriangle$}-}) and a least squares fitting method ({\large \color{red}{-$\bullet$-}}). The solid line represents white noise ($\tau^{-1/2}$ behavior) while the dashed line represents a random walk ($\tau^{+1/2}$) and the dashdotted line represents long-term drifts ($\tau^{+1}$ behavior). The lines between the points are merely a guide for the eye.}
\label{fig:DSP_vs_Fit}
\end{figure}
\indent We also compared frequencies measured by the DSP system with an offline numerical analysis using an exponentially decaying sinusoidal fit function. 
Figure~\ref{fig:DSP_vs_Fit} shows the performance of the DSP and the offline analysis for a duration of 1\,h.
At short $\tau$, the offline numerical analysis shows a better performance. 
As $\tau$ becomes larger than 20~s, both methods show the same result. 
At large values of $\tau$, where the offline mode of analysis takes a long time to compute $f_L$ for all data points, the DSP system is much more efficient. 
Figure~\ref{fig:DSP_vs_Fit} also shows a sensitivity of 700 \fth.
This is about 3.5 times larger than $\sigma_\mathrm{SN}$. 
By reducing environmental noise one expects to reach $\sigma_\mathrm{SN}$.\\
\indent Note that the DSP system is fast to track frequency variations, but its performance is limited to the specific frequency range defined by the bandwidth of its band-pass filter. In our case it is 3500\,$\pm$\,200\,Hz. 
Beyond that range, the DSP system will lose frequency tracking. 
However, one can still use our DSP to monitor the frequency of demodulated signals. 
In this way, any high frequency signals can be demodulated first and then sent to the DSP to extract their frequencies. 
In order to do that, a lock-in amplifier with a wide bandwidth and with a stable local oscillator oscillating at a frequency 3.5\,kHz away from the input signal is required. 
The output of the lock-in is a demodulated signal oscillating at 3.5\,kHz and can be used as the input of our DSP system.
In our case, the lock-in amplifier should have four input channels with preferably four independent local oscillators, one for each channel. 
We used this technique to actively measure the precession frequency of the FID signals oscillating at 35\,kHz when a 20\,mA current drives the field coil.
Such a DSP system is a promising solution for experiments requiring active and fast analysis of data from multiple sources. 
It has been used in precision measurements like the operation of an array of Cs magnetometers in the PSI nEDM experiment \cite{abel2020optically} or in magnetocardiography \cite{bison2009room, lembke2014optical}.  
The DSP system described here also served as a prototype for a larger system since it scales to support additional channels by using an SoC with more resources. 
The larger system features a 128 channel DSP, using two XCZU9EG boards, which will be used in the next generation of the nEDM experiment at the Paul Scherrer Institute, called n2EDM\cite{abel2019n2edm}.
\subsection{Magnetic Shields}
\indent For most measurements with Cs magnetometers, we used a 4-layer cylindrical mu-metal shield to isolate the magnetometers from external electromagnetic interference. 
The cylinders are closed on one end and open on the other end. Each open end has a removable end cap. 
All layers of the mu-metal shield were manufactured from ASTM~A753-08~alloy~type~4 with a thickness of 1.6~mm and were heat treated for maximum magnetic permeability. 
The layers of the shield are evenly spaced and the innermost layer is 80~cm long with a diameter of 42~cm. 
The outermost one is 92~cm long with a diameter of 54~cm. 
For measurements using K magnetometers we used the magnetic shield of the nEDM experiment at PSI~\cite{pendlebury2015revised}.\\
\section{Results}
We demonstrate the performance of the system for screening external magnetic interference and for tracking current variations. 
We also show that this system can detect small modulated currents under unshielded and shielded measurement conditions. 
Finally, we demonstrate a stabilized current source based on using the two-dimensional gradiometer configuration of four Cs magnetometers and a current feedback control system. 
\subsection{Magnetic Interference Screening}
The performance of the system in the presence of an external magnetic interference was evaluated by exposing the coil to the environmental laboratory conditions, i.e., by removing it from the mu-metal shield. 
As the system operates at low magnetic fields, we used three pairs of square Helmholtz coils to compensate the projection of the static component of the background magnetic field along the measurement axes of the coil. 
Then we evaluated how well \textbf{B}$_\mathrm{ext}$ is detected and cancelled by the system. 
We tested the system at driving currents larger than 10 mA. 
The system can operate at lower currents if the background field reduces to a level lower than the field generated by the coil.\\
\indent The background magnetic field gradients degrade the quality of the FSP signal and increase its decay rate. 
This in turn lets us increase the sampling rate in current measurements up to 50 S/s.
\begin{figure}[t!]
	\includegraphics[width=0.5
\textwidth]{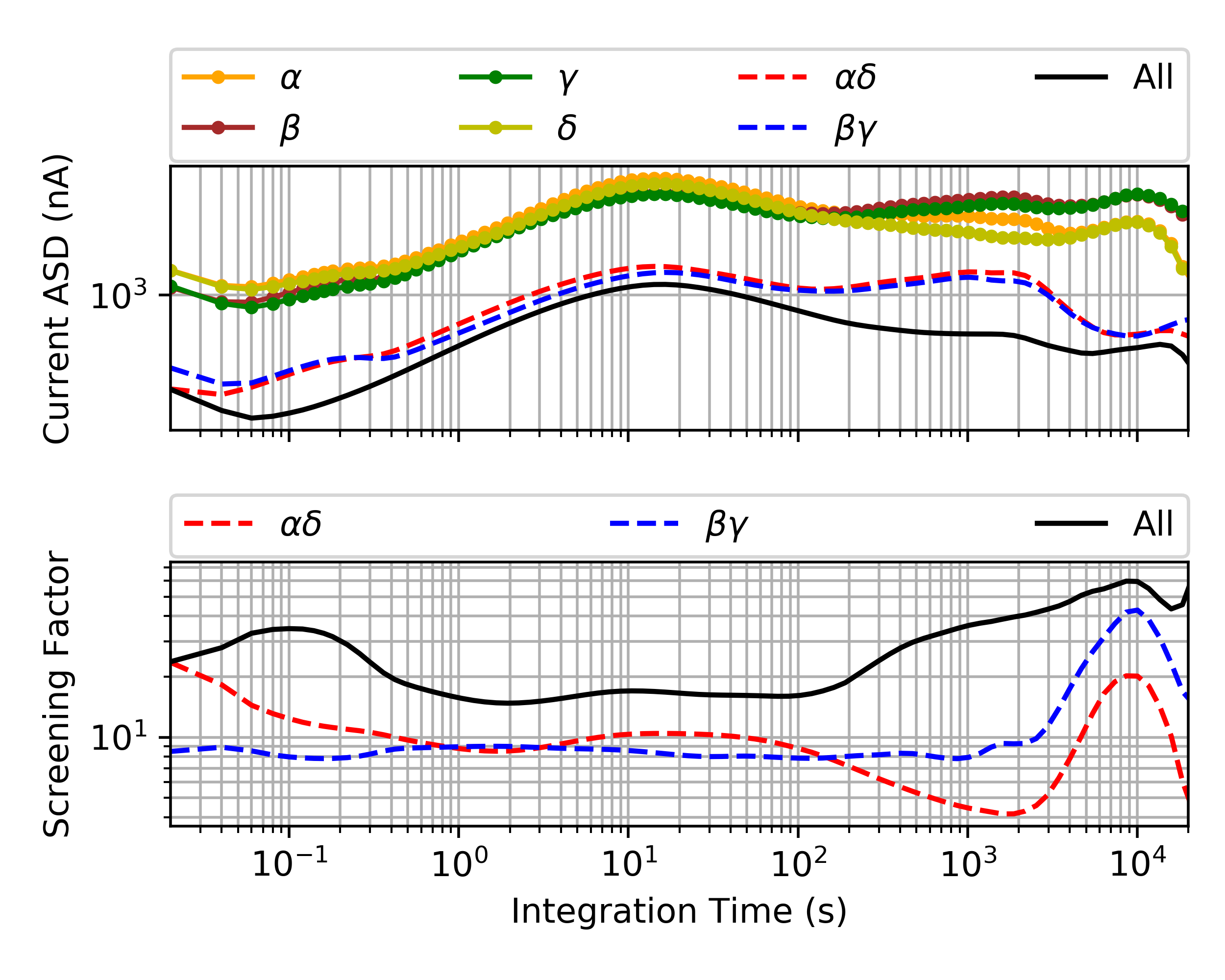}
	\caption{Top: ASD plot of the measured current of the coil using Cs magnetometers at room temperature as a function of integration time. Bottom: Corresponding screening factor of the gradiometer configuration of our system. The coil stood in an unshielded region and it was driven by a 12 mA current. The different traces represent data for different quadrants and combinations of quadrants.}
	\label{fig:unshielded_performance}
\end{figure}
Figure~\ref{fig:unshielded_performance} shows the ASD of the measured driving current of the coil and the screening factor associated with the gradiometer configuration of the coil in an unshielded environment. 
For this measurement, the coil was driven with a 12 mA current. 
The screening factor shows to which extent the ASD of  the current averaged over all or some of modules reduces compared to the ASD of the current obtained from a single module. 
From Fig.~\ref{fig:unshielded_performance} we see that the screening associated with all modules outperforms those associated with modules $\alpha \delta$ and $\beta \gamma$. 
At short integration times ($\tau$\,=\,100\,ms) the screening factor reaches a value of 34. 
This factor can be as high as 60 for very long integration times.  
Note that the performance of the system in screening depends on the uniformity of \textbf{B}$_\mathrm{ext}$, the design of the coil and the orientation of modules in respect to \textbf{B}$_\mathrm{ext}$. 
Nevertheless, one expects that the system stability improves by more than one order of magnitude due to the screening effect of the gradiometer configuration.
\subsection{Current Tracking}
\begin{figure}[t!]
	\includegraphics[width=0.5
\textwidth]{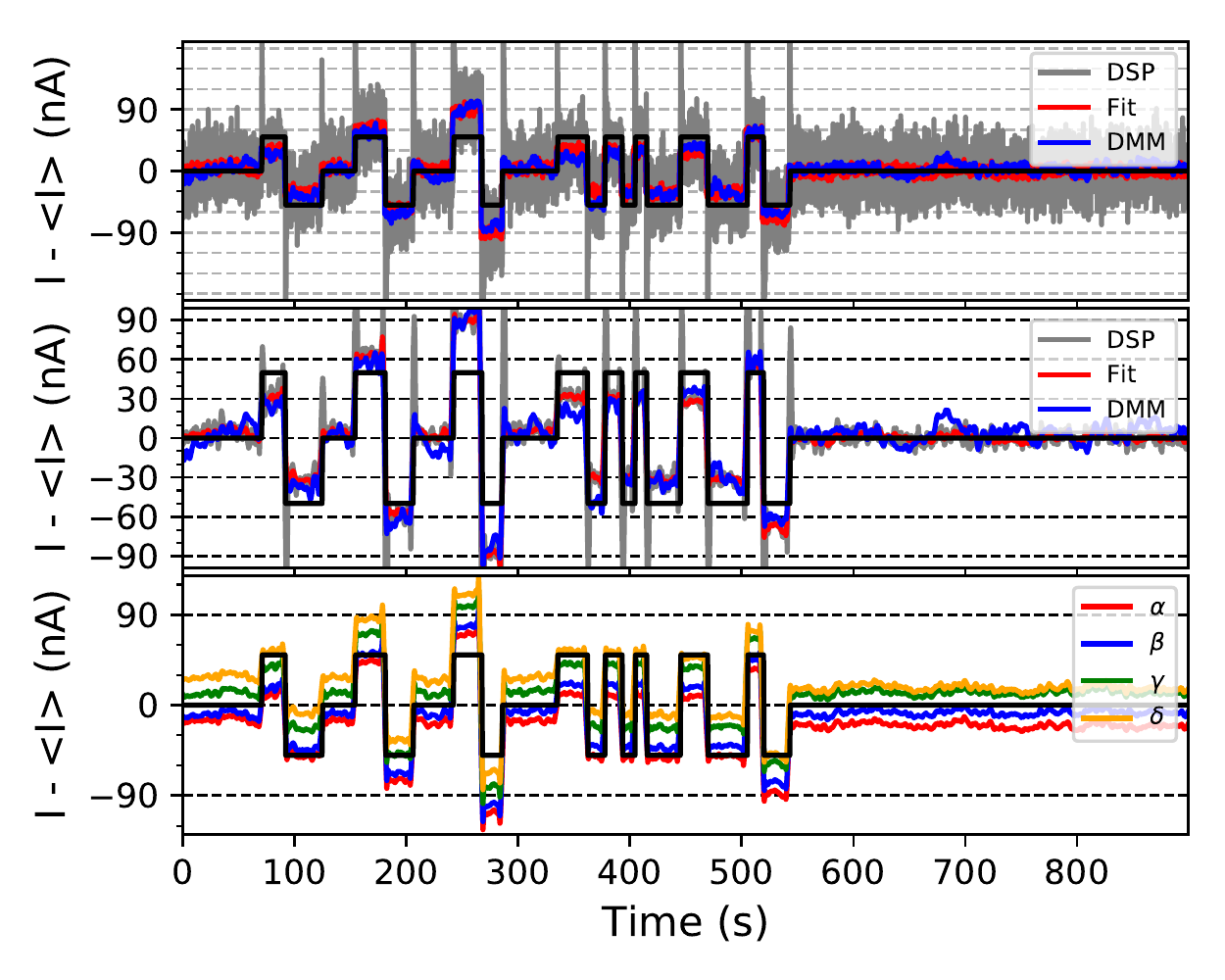}
	\caption{Response of the system and the DMM to sudden changes in current. The system's data is analyzed using digital-signal processing (DSP) and least squares fitting (Fit). The top panel shows the system and DMM responses without filtering but the middle panel shows filtered responses using a 3rd order Savitzky-Golay filter with a window length of 5~sec. The bottom panel shows individual responses of all four Cs magnetometers ($\alpha, \beta, \gamma$ and $\delta$) filtered using the same method as the middle panel. Individual responses are shifted by small offsets to demonstrate their correlation. The black solid line represents the trigger signal where the current in the coil is changed by discrete steps of $\pm$30, $\pm$60 and $\pm$90~nA. The grey spikes show where the PID of our DSP tries to find the right values after a sudden change in the coil current.}
	\label{fig:CMS_and_step_changes}
\end{figure}

Figure \ref{fig:CMS_and_step_changes} shows the response of the system to $\pm$30, $\pm$60 and $\pm$90~nA step changes in current passing through the coil. 
We used the DSP and least-square fitting methods to analyse currents sensed by the system. 
The power of the beams of all four magnetometers was set to the region of optimal sensitivity as shown in Fig.~\ref{fig:Cs_K_sens}. 
The magnetometers were placed in the coil at a 1 \mt~magnetic field, which was produced by a 2~mA driving current. 
The figure also shows how well all magnetometers can track each other in the presence of sudden changes. 
We found a correlation better than 99.4$\%$ between each channel and the average of all four channels.   
The performance of our system was compared with a 7$\frac{1}{2}$ digit digital multimeter (DMM) with a sampling rate of 0.5 s$^{-1}$ and an integration time of ten 50\,Hz power line cycles. 
The DMM was used to monitor voltage drops across a 100\,$\Omega$ resistor with a 0.2\,ppm/\cels~temperature coefficient. 
The correlation between the DMM and the system was measured to be 92$\%$. 
One can notice that our system outperforms the DMM as it shows less deviation from expected values of the current.  
\subsection{Modulated Current}
\begin{figure}[t!]
	\includegraphics[width=0.5
\textwidth]{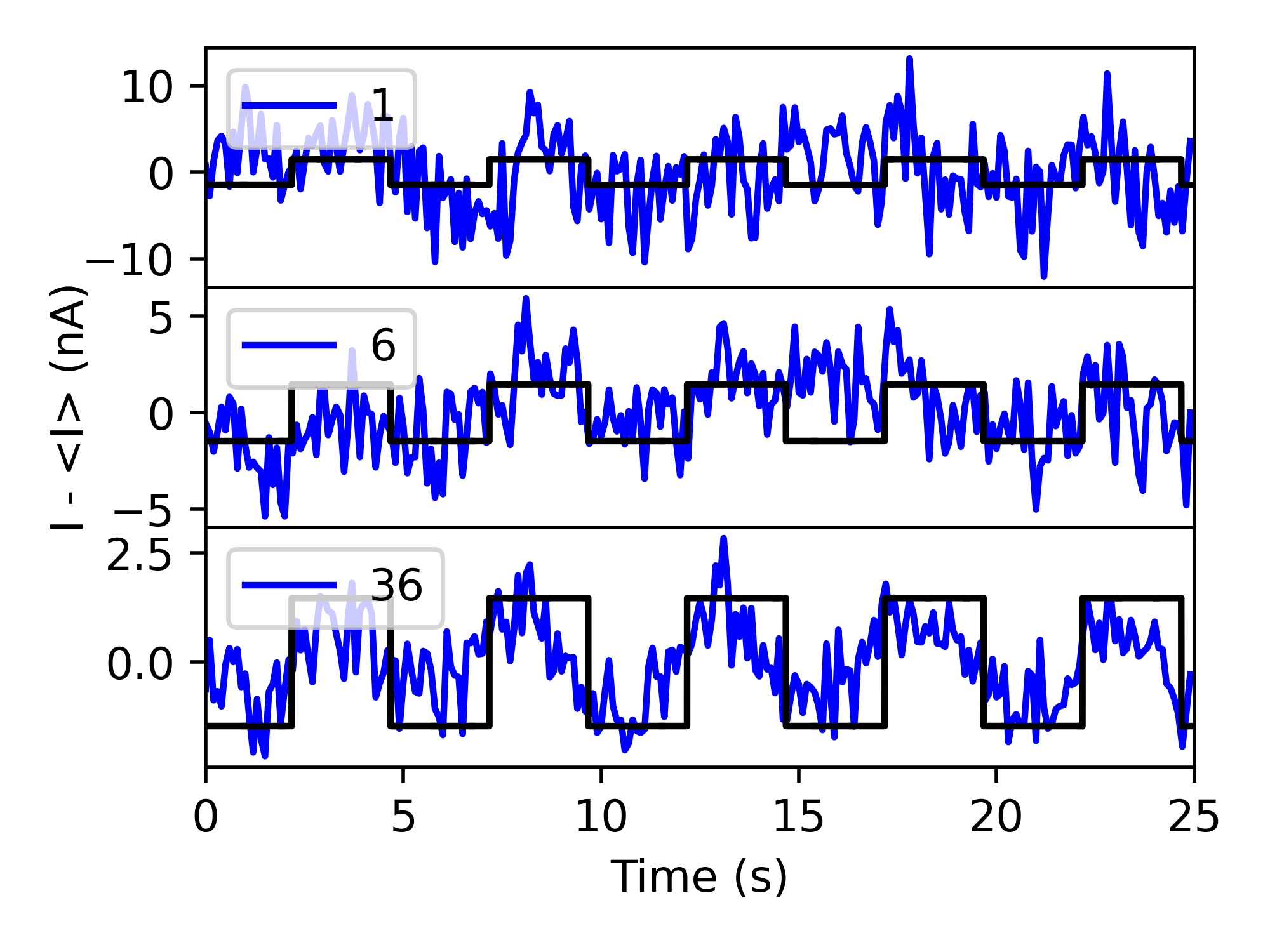}
	\caption{Modulated current signals with an amplitude of 2\,nA acquired at $f_{mod}$~=~0.2~Hz . Three examples are shown. The top panel shows a single shot acquisition. The middle
panel shows the effect of averaging over 6 repetitions of this experiment and the bottom
panel shows the effect of averaging over 36 repetitions. The square modulation pattern
used to turn the nA current on and off is also shown.}
	\label{fig:frequency_responce_modulated}
\end{figure}

\begin{figure}[t!]
	\includegraphics[width=0.5
\textwidth]{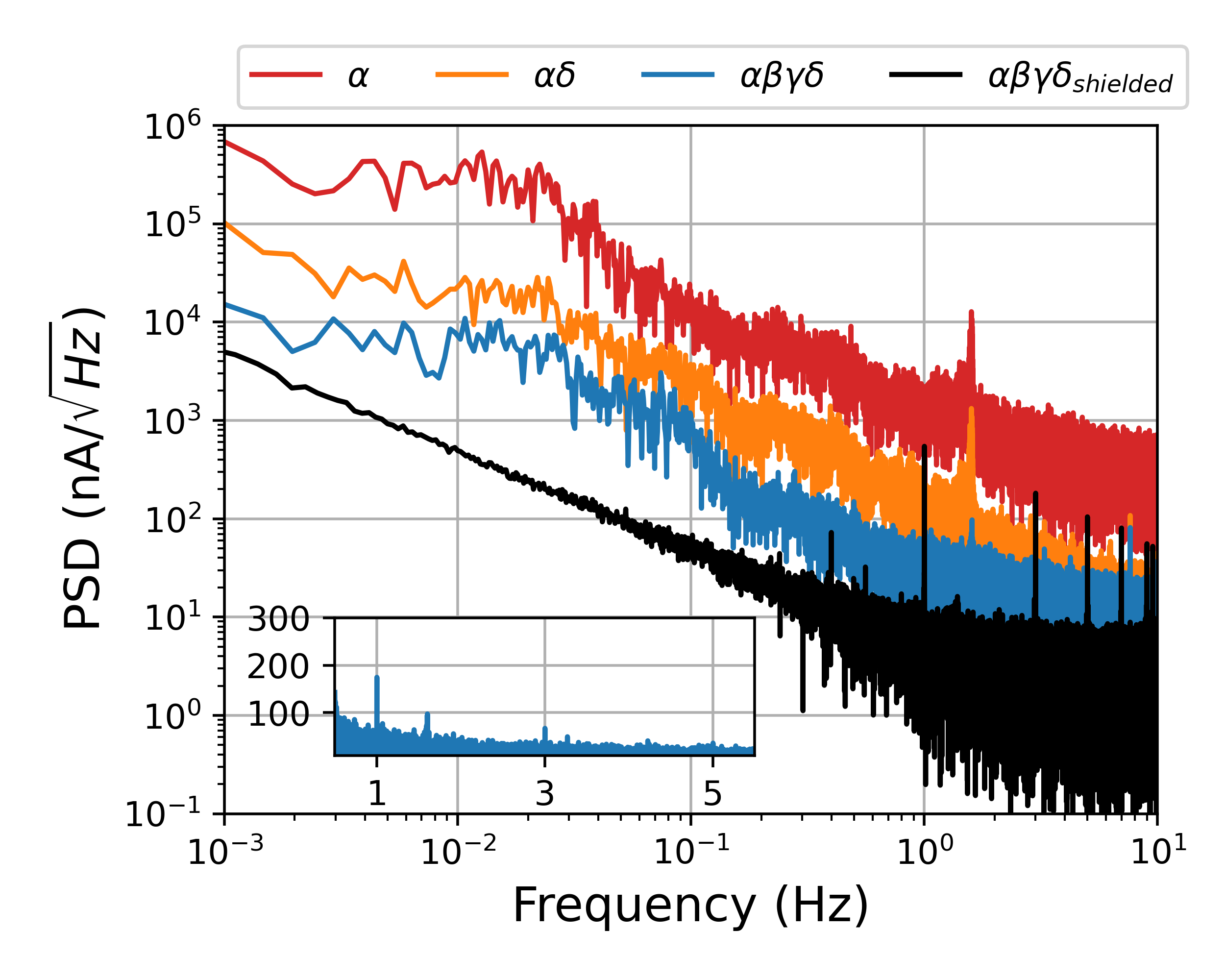}
	\caption{Power spectral density (PSD) of the measured current at 12~mA in an unshielded environment after 1\,h of data acquisition when a 20\,nA modulated current at $f_{mod}$~=~1~Hz was added to the 12\,mA current. The current was measured using module $\alpha$ (red), $\alpha \delta$ (orange) and all four modules (blue). The black color data shows the performance of the system in a similar measurement using all four modules but in a shielded environment. The spikes at 1\,Hz and its odd multiples are associated with $f_{mod}=$1\,Hz and higher harmonics. Inset: Expanded view of the PSD for all modules in the unshielded environment on a linear plot. }
	\label{fig:PSD}
\end{figure}

\indent As a test of the viability of the optical magnetometers to monitor small variations in current, we modulated the driving current with a modulation frequency $f_{mod}$.
This imprints an alternating pattern on the magnetic field inside the coil. 
The optical magnetometers were then used to directly monitor the resulting magnetic field variations and consequently current variations. 
$f_{mod}$ is limited by the sampling rate of the system. 
In an unshielded region, where the field non-uniformity is high, the FSP signals decay rapidly. 
This results in a sampling rate of up to 50\,S/s. 
In a very well shielded region, the typical sampling rate is 10\,S/s. 
To achieve a higher sampling rate we have increased the pump/probe laser powers in order to shorten the FSP decay rate. 
This led us to acquire a sampling rate of 50\,S/s. 
At higher sampling rates, the DSP fails to process the FSP signals, as it typically requires 30\,ms to process them.\\  
\indent Figure\,\ref{fig:frequency_responce_modulated} shows recordings of the averaged amplitude of the modulated current, sensed by all magnetometers, when the coil is driven by a 2\,mA current. 
The coil stood in a shielded region with 4 layers of mu-metal.
The modulation pattern is a square waveform with a frequency $f_{mod}$\,=\,0.2\,Hz that activates a mechanical relay with a delay time of 100\,ms. 
The relay connects/disconnects a secondary current source to/from the main current source. 
The former generates nA currents and the later drives the coil with mA currents. 
The figure shows how a small amplitude of a modulated current can be monitored by the system in a shielded environment using all four modules.\\
\indent To compare the performance of our system in an unshielded and well shielded region, we calculated the power spectral density (PSD) of the measured current of the coil which was the sum of a 12\,mA DC and a 20\,nA current modulated at 1\,Hz. 
We used a standard periodogram method to calculate the PSD. 
Figure\,\ref{fig:PSD} shows results of the PSD after 1\,h of data acquisition. 
The data shows how the PSD is improved when we go from using a single module ($\alpha$), to two ($\alpha \delta$) and four ($\alpha \beta \gamma \delta$) modules. 
Two and four modules correspond to one-dimensional and two-dimensional gradiometer configurations respectively. 
It also demonstrates how the two-dimensional configuration in a noisy region, without using any magnetic shield, can resolve the modulated signal. This signal appears as a series of spikes at $f_{mod}$\,=\,1\,Hz and its higher harmonics up to the 5th order. 
The signal to noise ratio (SNR) at $f_{mod}$ = 1\,Hz is about 5. 
By using a 4-layer mu-metal shield, we see much improvement in the PSD as expected. 
It results in a SNR $\sim$ 95. 

\subsection{Current Feedback Control}
\begin{figure}[t!]
	\includegraphics[width=0.4
\textwidth]{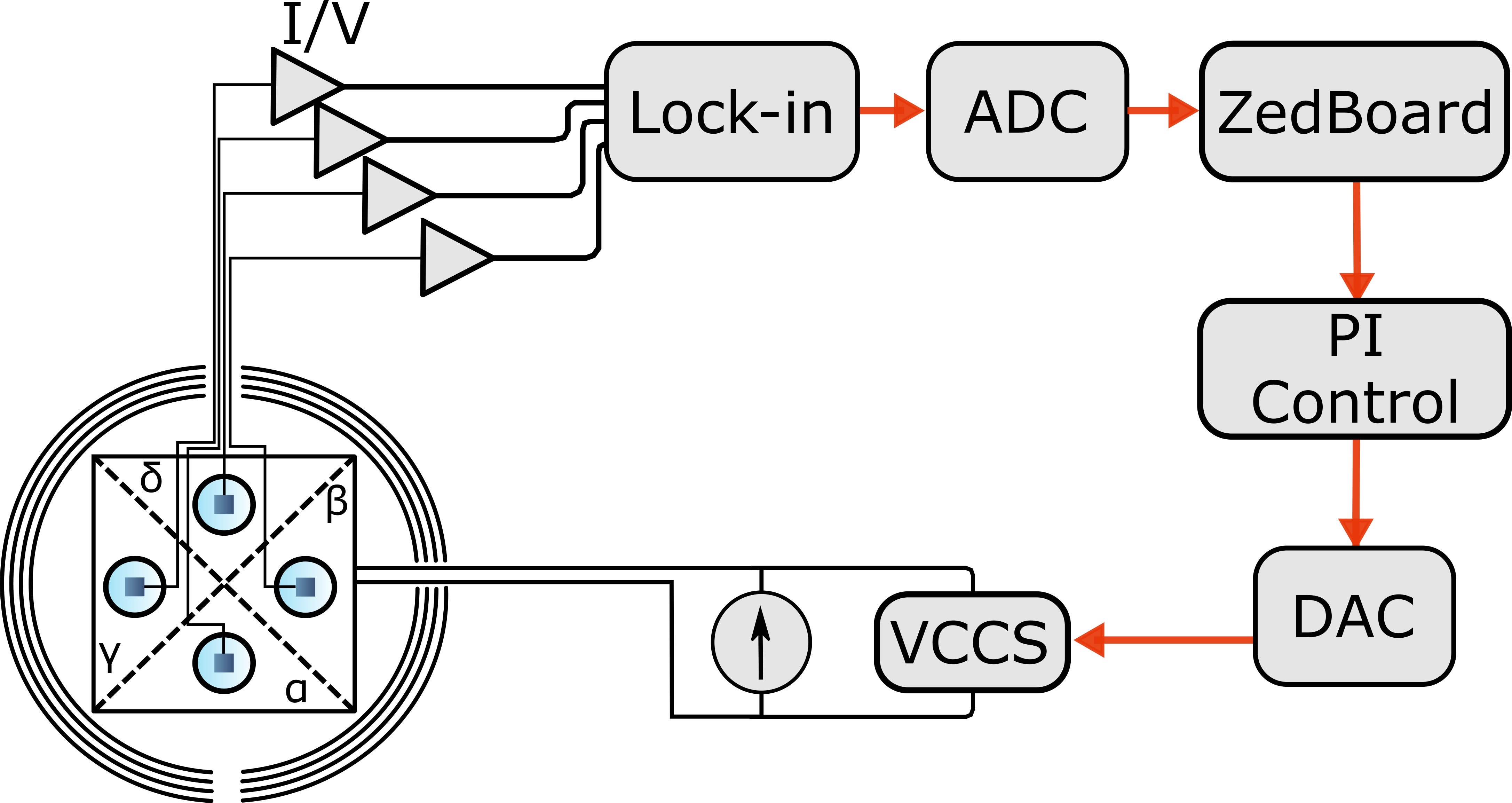}
	\caption{Signal path for a current feedback control loop. The amplified photocurrent signals of all four modules are demodulated by a four input channel lock-in amplifier with four independent local oscillators. The demodulated output signals of the lock-in are sent to the DSP to measure the driving current of the coil. The output of the Proportional-Integral (PI) control converts to an analog signal using a digital-to-analog converter (DAC) and then is sent to a voltage control current source (VCCS). The VCCS output is added to the output of the main current source to compensate coil's current drift.}
	\label{fig:feedback_scheme}

\end{figure}

\begin{figure}[t!]
	\includegraphics[width=0.5
\textwidth]{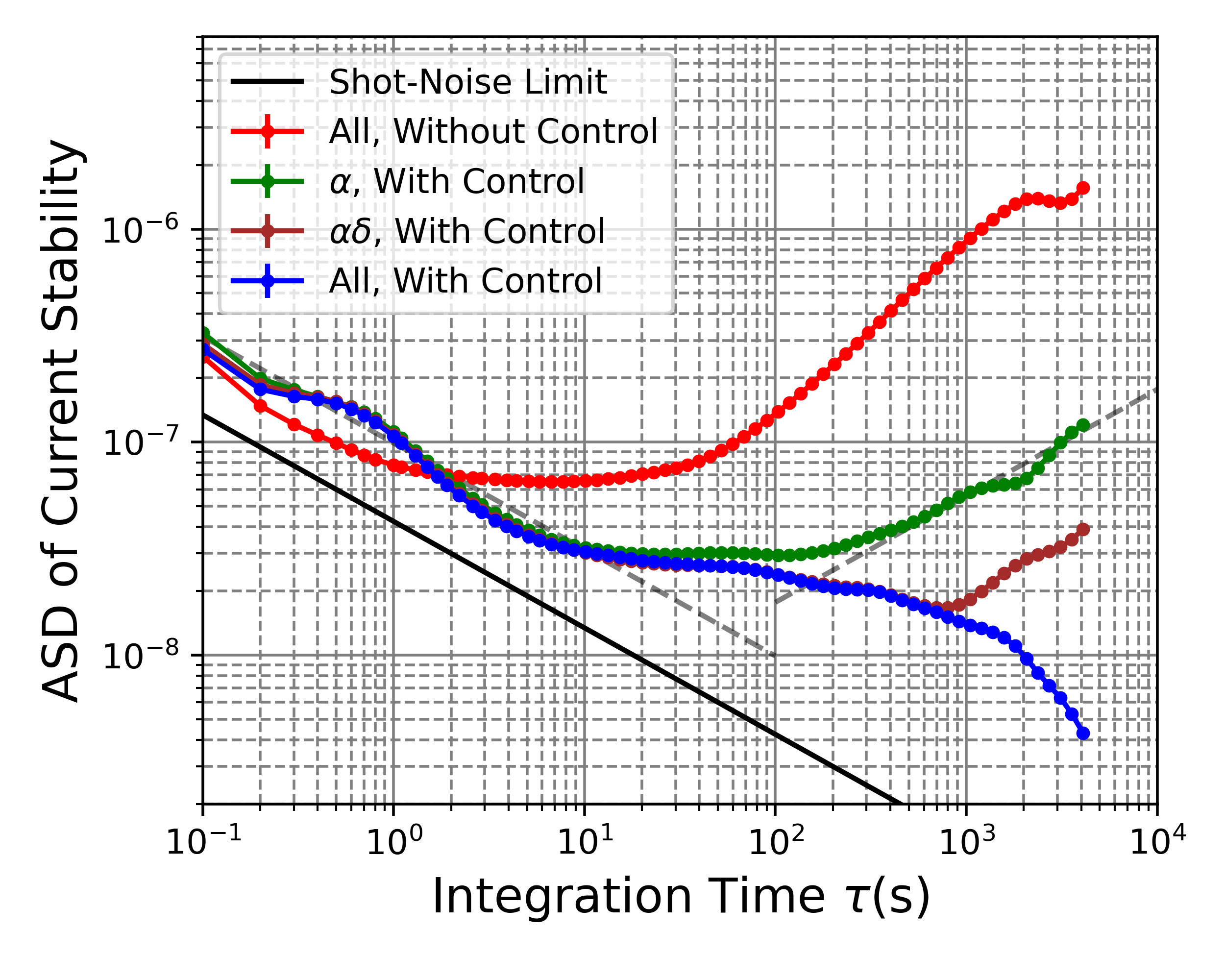}
	\caption{ASD plot of the current stability at 20\,mA, which corresponds to a magnetic field of 10\,\mt~inside the field coil. Cs magnetometers were used to monitor the coil's current when no current feedback control was implemented (red), when the current feedback control was applied but the current was measured using a single (green), two (brown) and all four (blue) magnetometers. The dashed lines represent white noise (left) with $\tau^{-1/2}$ and random walk behavior (right) with $\tau^{+1/2}$. The solid line shows the best achievable performance based on the shot noise limited sensitivity of the Cs magnetometers.}
	\label{fig:ASD_feedback}
	% data from February 15, Ti = 0.1, 75 ms data used for fitting, 60 ms used for FPGA, PID parameters: -900, 0.1*60 s and 0 for Kc, Ti and Td respectively. I used the first 25000 sec of data.
\end{figure}

The current applied to the field coil stabilizes through a feedback control scheme, which is shown in Fig. \ref{fig:feedback_scheme}. 
In this scheme, the amplified photocurrent signals are first demodulated using a four input channel lock-in amplifier with four independent local oscillators. 
As our DSP system can track signals at $f_{DSP}$\,=\,3.5\,kHz, we use a lock-in to demodulate the FSP signals oscillating at high frequencies such that the demodulated signals have a frequency close to $f_{DSP}$.
Thus, signals at higher frequencies are trackable by our DSP system.
In order to achieve that, the lock-in bandwidth should be wide enough and the frequency of the local oscillators must be shifted from the frequency of the FSP signals by $f_{DSP}$. 
The typical bandwidth of the lock-in used in our measurements was 4.5\,kHz. 
In the particular case where the current of the coil is 2\,mA the lock-in can be bypassed.
The demodulated outputs of the lock-in are first digitized.
Then they are sent to the DSP system to extract their frequencies. 
The DSP calculates $\bar{f}$ the average of these frequencies and $\delta f$ the deviation of $\bar{f}$ from $f_{DSP}$.
$\delta f$ is used as the error signal for the current feedback control that consists of a Proportional-Integral (PI) control algorithm, a digital-to-analog converter (DAC) and a voltage controlled current source (VCCS) with a typical gain of 100\,nA/V. 
The VCCS was used to compensate current fluctuations of the main current source.\\
\indent Figure \ref{fig:ASD_feedback} shows the ASD plot of the current stability when the drift in the coil's current was stabilized using the current feedback control. 
The coil's current was 20\,mA which corresponds to a magnetic field of 10 \mt~inside of the coil. 
We used the offline numerical analysis rather than the online method to calculate the frequencies of FSP signals and the driving current correspond to them. 
The reason to use offline method is that it outperforms the online method at short integration times.  
For comparison, we showed the current stability at 20\,mA with and without current feedback control. The data associate with each case was recorded at different times but under the same experimental conditions. 
As Fig.\,(\ref{fig:ASD_feedback}) shows, using a feedback control results in significant improvements of the current stability, especially on the long time scales.
The best stability we could achieve was 4\,$\times$\,10$^{-9}$ after 70\,min of averaging. 
The figure also shows how the stability improves by extending the gradiometer feature from one pair to two pairs of magnetometers. 
The current stability based on using a single ($\alpha$) and a pair ($\alpha \delta$, one-dimensional gradiometer) of Cs magnetometers start drifting after 200 and 800 seconds.
However, using two pairs of Cs magnetometers, which corresponds to a two-dimensional gradiometer configuration, keeps improving even after an hour of averaging. This shows how well two-dimensional gradiometer configuration can screen the effect of external magnetic interference.
The technical implementation of this system could still be slightly improved, as the sensitivity of the magnetometers is worse than the shot noise limited sensitivity. By suppressing technical noises and operate with shot noise limited sensitivity one can achieve 4\,$\times$\,10$^{-9}$ just after 100\,sec averaging. 
\section{Outlook}
\begin{figure}[t!]
    \subfloat[\label{subfig:triangle}]{%
    \includegraphics[width=0.5\columnwidth]{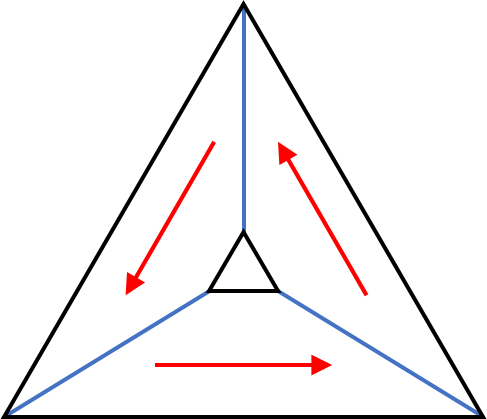}%
    }\hspace{0.05\columnwidth}
    \subfloat[\label{subfig:square}]{%
    \includegraphics[width=0.4\columnwidth]{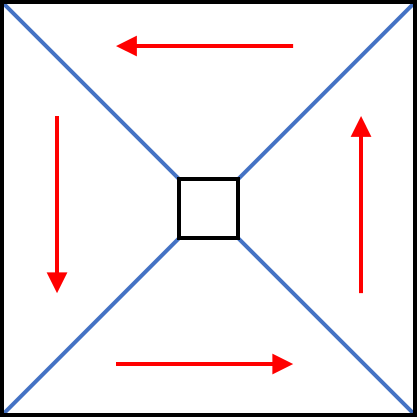}%
    }\\
    \subfloat[\label{subfig:hexagon}]{%
    \includegraphics[width=0.5\columnwidth]{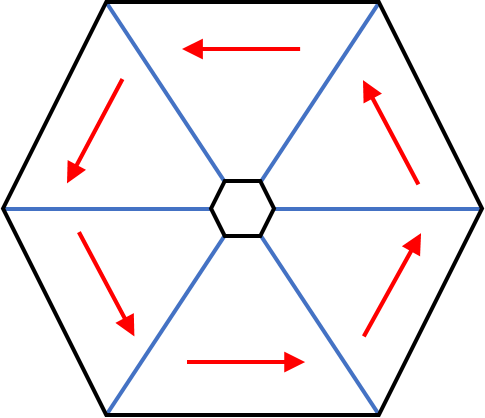}%
    }\hspace{0.05\columnwidth}
    \subfloat[\label{subfig:composed_square}]{%
    \includegraphics[width=0.4\columnwidth]{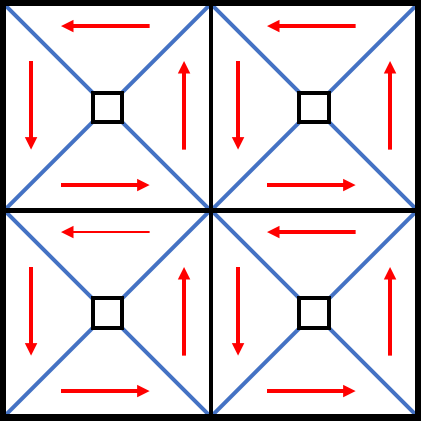}%
    }
	\caption{Schematic representation of a triangular (a), a square (b) and a hexagonal (c) coil. The black boundaries represent a zero flux condition, i.e., no field can pass. The blue boundaries represent the flux conditions to guide the magnetic field from one section to the next. The red arrows represent the magnetic field direction in each section. The PCB design of these coils allows for more complex wire paths and more complex coil systems to be built. These simple coils can be used as building blocks for more complex constructs as shown on (d).}
	\label{fig:lego_coils}
\end{figure}

The system presented in this paper was developed for the nEDM experiment at the Paul Scherrer Institute.
However, several new developments in the field of OPM could improve the accessibility and applicability of this technology.
The last years have shown the appearance of startup companies, which specialize in the commercialization of OPM.
The focus of these companies is to make use of microfabricated vapor cells, which can be used to make chip-scale OPM~\cite{schwindt2004chip}.
To the best of our knowledge, three companies are currently selling such sensors: QuSpin, Twinleaf and FieldLine.
Of particular interest to the system presented in this paper is the new gradiometer sold by Twinleaf~\cite{limes2020portable}.
This sensor could be used in an array similar to the one presented here.
This would simplify the access to this technology.
Furthermore, the research in the field of OPM is seeing a clear shift towards more robust sensors, which can be used in a variety of environments~\cite{fu2020sensitive}.
The bandwidth of OPM can also be improved significantly, which would allow the system to be used with AC currents of a wide frequency range~\cite{wilson2020wide}.\\
\indent The practical benefits of this system may seem limited due to the dependence on the relative sensitivity.
However, much of the negative sides can be mitigated or significantly suppressed. 
The current at which this system operates can be changed simply by changing the coil constant.
Thus, the optimal operating field for the sensors can be chosen for the current one wants to stabilize.
Compared to a DMM the range of operation is limited since the relative performance degrades due to gradients and non-linearities.
However, the continued progress and miniaturization of OPM should significantly mitigate this problem~\cite{fu2020sensitive, schwindt2004chip}.\\
\indent Figure~\ref{fig:lego_coils} shows several variations of the cubic coil geometry used in this work.
The design method for this type of coil is based on the magnetic scalar potential $\Phi_M$.
It allows us to define the desired magnetic field and compute the location of the wires required to generate this magnetic field.
For simple geometric shapes like in Fig.~\ref{fig:lego_coils} the wires will always be equidistant and evenly spaced like for the coil prototype of Fig.~\ref{fig:V2}~\cite{koss2017pcb}.
Having more sensors at different locations will improve the discrimination feature of this system. 
Especially a design like the one shown in Fig.~\ref{subfig:composed_square} would allow to discriminate and even measure higher order gradients.
The PCB design of the coil would allow the use of card edge connectors, which makes the construction of such a coil feasible.
\section{Conclusion}
\indent We presented a current monitoring system, whose concept can be adapted to detect current variations of any current source. 
It features a field-confining coil, which contains four optically pumped magnetometers.
This allows us to exploit the sensitivity of these magnetometers and convert the magnetic sensitivity into a current sensitivity.
When the setup is well shielded and thermally stable, the sensitivity of this system is mainly limited by the sensitivity of the magnetometers. 
For Cs and K magnetometers operating in a free spin precession mode in a 1 \mt~holding field we measured a shot noise limited sensitivity of about 200 \fth~and 20 \fth~respectively.
This corresponds to a current sensitivity of 400 \pah~and 40 \pah~respectively. 
The best performance was observed for magnetic fields larger than 1 \mt~and using Cs magnetometers.
A relative sensitivity of better than $5 \times 10^{-9}$ was achieved when the coil magnetic field is 10 \mt~in an actively controlled setup. 
For this setup, this corresponded to a 20 mA driving current, which has a stability on the 100 pA level. 
The stability of this system can still be improved.
For example by improving the sensitivity of the magnetometers or by using a coil with a higher coil constant to increase the relative sensitivity.\\
\indent It was shown that this current monitoring system is able to track changes in current. 
It can also discriminate external magnetic perturbations through the first order gradiometer configuration in which the magnetometers arrange. 
The performance of the gradiometer can be improved by reducing the distance between magnetometers or by adding more sensors to the system.
Finally, implementing a current feedback control stabilizes a current source and achieves a high sensitivity in magnetic field and current sensing over long integration times. \\ 
\begin{acknowledgments}
This work was partly supported by the Fund for Scientific Research Flanders (FWO) and Project No. GOA/2010/10 of the KU Leuven.
The authors would like to thank the electronics workshop of the KU Leuven physics department for designing the PCB and constructive feedback on the design of the coil used in this work.
\end{acknowledgments}

%\nocite{*}

\bibliography{monitor_paper}% Produces the bibliography via BibTeX.

\end{document}